\documentclass[%
 reprint,
superscriptaddress,
 amsmath,amssymb,
 aps,
 prb,
 floatfix,
]{revtex4-2}
\usepackage{CJK}
\usepackage{amsmath}
\usepackage{array}
\usepackage{dsfont}
\usepackage{mathtools}
\usepackage{graphicx}
\usepackage{dcolumn}
\usepackage{bm}
\usepackage[pdfencoding=auto]{hyperref}
\usepackage{physics}
\usepackage{empheq}
\usepackage{color}
\usepackage{multirow}
\usepackage{afterpage}
\usepackage{mathrsfs}
\usepackage{tikz}
\CJKnospace

\usetikzlibrary{shapes}
\usetikzlibrary{tikzmark,calc}

\newcommand{\hexagon}{\mathord{\raisebox{0.6pt}{\tikz{\node[draw,scale=.65,regular polygon, regular polygon sides=6](){};}}}}

\begin{document}
\begin{CJK*}{UTF8}{gbsn}
\title{Magnetic Field Response of Dipolar-Octupolar Quantum Spin Ice}

\author{Zhengbang Zhou (周政邦)}
\email{zhengbang.zhou@mail.utoronto.ca}
\affiliation{%
 Department of Physics, University of Toronto, Toronto, Ontario M5S 1A7, Canada
}%
\author{F\'elix Desrochers}
\email{felix.desrochers@mail.utoronto.ca}
\affiliation{%
 Department of Physics, University of Toronto, Toronto, Ontario M5S 1A7, Canada
}%
\author{Yong Baek Kim}%
\email{ybkim@physics.utoronto.ca}
\affiliation{%
 Department of Physics, University of Toronto, Toronto, Ontario M5S 1A7, Canada
}%

\date{\today}
\begin{abstract}
Dipolar-octupolar (DO) pyrochlore systems Ce$_2$(Zr,Sn,Hf)$_2$O$_7$ have garnered much attention as recent investigations suggest that they may stabilize a novel quantum spin ice (QSI), a quantum spin liquid (QSL) with an emergent $U(1)$ gauge field. In particular, the experimentally estimated microscopic exchange parameters place Ce$_2$Zr$_2$O$_7$ in the $\pi$-flux QSI regime, and recent neutron scattering experiments have corroborated some key theoretical predictions. On the other hand, to make a definitive conclusion, more multifaceted experimental signatures are desirable. In this regard, recent neutron scattering investigation of the magnetic field dependence of the spin correlations in Ce$_2$Zr$_2$O$_7$ may provide valuable information. However, there have not been any comprehensive theoretical studies for comparison. In this work, we provide such information using gauge mean-field theory (GMFT), allowing for theoretical investigation beyond the perturbative regime. In particular, we construct the phase diagrams for the [110], [111], and [001] field directions. Furthermore, we demonstrate the distinctive evolution of the equal-time and dynamical spin structure factors as a function of the magnetic field for each field direction. These predictions will help future experiments confirm the true nature of the DO-QSI.
\end{abstract} 
\maketitle
\end{CJK*}

\section{\label{sec: Introduction} Introduction}

Quantum spin liquids (QSLs) are paramagnetic quantum ground states of frustrated spin systems exhibiting long-range entanglement (LRE)~\cite{wen2002quantum, wen2004quantum, wen2013topological, balents2010spin, savary2016quantumspinliquids, knolle2019field} that host fractionalized excitations along with emergent gauge fields. Because of the lack of long-range order (LRO) and the corresponding absence of order parameters, the experimental identification of a QSL has proved challenging and is still an ongoing endeavor in condensed matter physics~\cite{wen2019experimental}. 

\begin{figure*}
    \centering
    \includegraphics[width=0.9\linewidth]{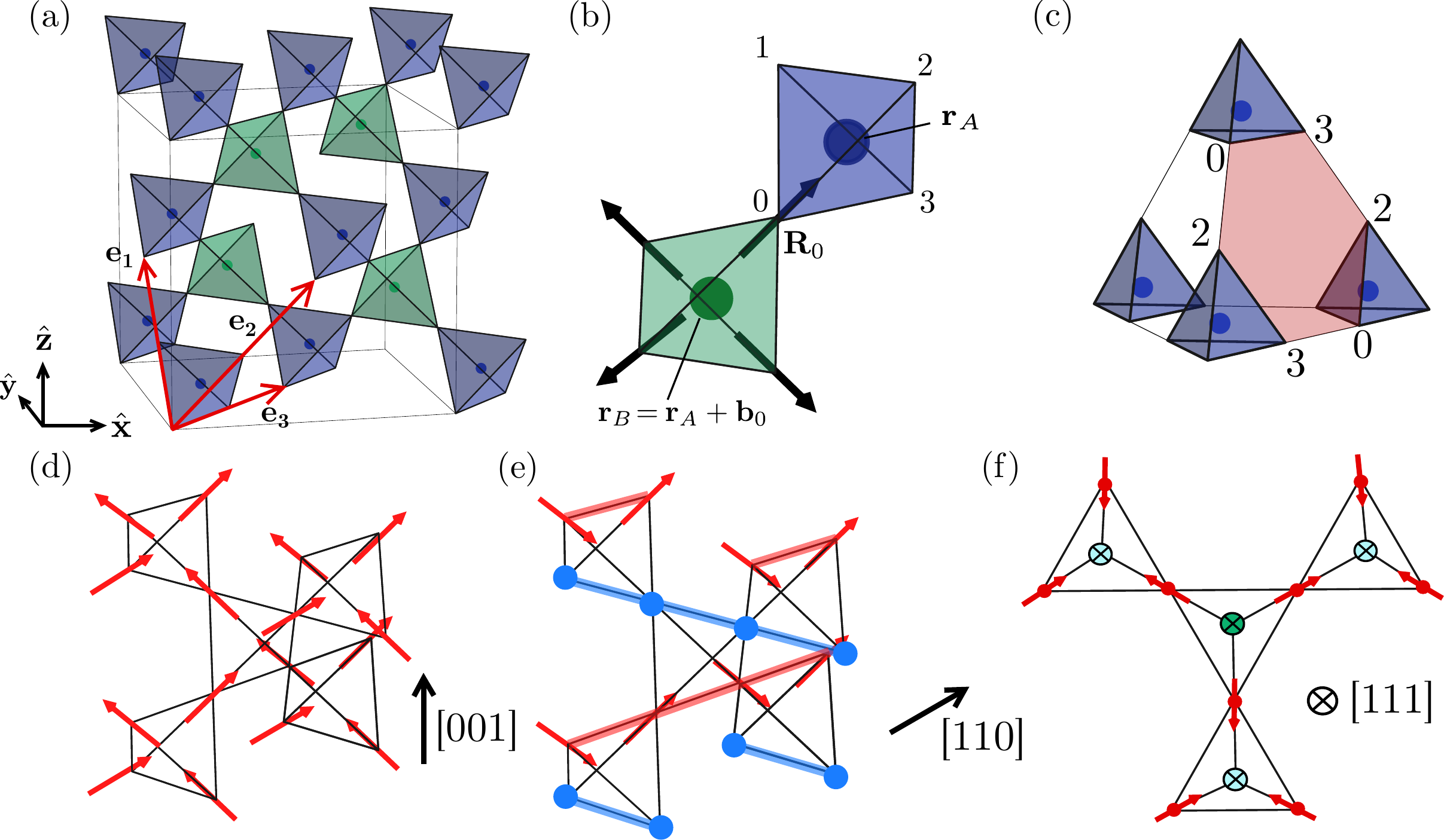}
    \caption{(a) The pyrochlore lattice. The down-pointing (up-pointing) tetrahedrons are colored blue (green). The parent diamond lattice sites located at the center of each tetrahedron are illustrated by spheres. (b) A single diamond unit cell with the four pyrochlore sublattices labeled. Black arrows denote the local pseudospin axis $\hat{\mathbf{z}}_\mu$ on each sublattice. The parent diamond lattice sites are labeled with $\mathbf{r}_A$ and $\mathbf{r}_B$. (c) Inequivalent hexagonal plaquettes. The plaquette $F_{320}$, which involves pyrochlore sublattices 0, 2, and 3, is highlighted in red. Polarized pseudospin $S^z$ configuration in the presence of only the Zeeman term (i.e., $J_{yy}/h\to 0$ and $J_{\pm}/h\to 0$) for a field along the (d) [001], (e) [110], and (f) [111] directions. The blue sites in (e) denote the $\beta$ chains that are decoupled from the field. For the [111] field in (f), the red sites form a Kagome plane, whereas the green sites make up triangular planes. Notice that the light and dark green sites are on different planes along the $[111]$ axis.}
    \label{fig:pyrochlore}
\end{figure*}

Extensive experimental efforts have been made in the search for QSLs~\cite{bramwell2020thehistory, sibille2018experimental}. In particular, recent works on Ce$_2$Zr$_2$O$_7$~\cite{gaudet2019quantum, gao2019experimental, smith2022case, gao2022magnetic, smith2023dipole, smith2023quantum, Beare2023muSR, gao2024emergent}, Ce$_2$Sn$_2$O$_7$~\cite{sibille2015candidate, yahne2022dipolar, sibille2020quantum, poree2023fractional}, and Ce$_2$Hf$_2$O$_7$~\cite{poree2023dipolar, poree2023dipolar2, bhardwaj2024thermodynamics} have been particularly promising. Measurements indicate that these so-called dipolar-octupolar (DO) compounds may host quantum spin ice (QSI), a QSL with a compact $U(1)$ emergent gauge structure that provides a realization of quantum electrodynamics (QED) on the pyrochlore lattice (Fig.~\ref{fig:pyrochlore}(a))~\cite{hermele2004pyrochlore, ross2011quantum, banerjee2008unusual, shannon2012quantum, kato2015numerical, gingras2014quantum, udagawa2021spin, chern2019magnetic, castelnovo2012spin}. QSI is theoretically predicted to host an emergent gapless photon mode as well as gapped spin-1/2 spinon excitations~\cite{benton2012seeing, rau2019frustrated, savary2021quantum, lee2012generic, benton2018quantum, taillefumier2017competing, huang2018dynamics}. Heat capacity and muon spin relaxation measurements have been conducted on these materials, and no sign of LRO or spin freezing has been observed down to the lowest accessible experimental temperature. The microscopic exchange parameters have been estimated by fitting various measurements using numerical linked cluster (NLC), Lanczos method, and semi-classical dynamics~\cite{bhardwaj2022sleuthing, smith2022case, smith2023quantum, poree2023dipolar2, bhardwaj2024thermodynamics, tang2013short}. All studies (except one on Ce$_2$Sn$_2$O$_7$~\cite{yahne2022dipolar}) place the materials in a region of parameter space that is theoretically predicted to host $\pi$-flux QSI~\cite{hermele2004pyrochlore, patri2020distinguishing, benton2020ground, chern2024pseudofermion}. In $\pi$-flux QSI, the hexagonal plaquettes of the pyrochlore lattice shown in Fig.~\ref{fig:pyrochlore}(c) are threaded by a static $\pi$-flux of the emergent gauge field.

There has also been more direct evidence for the possible realization of QSI in Ce$_2$(Zr,Sn,Hf)$_2$O$_7$. In Ce$_2$Zr$_2$O$_7$, the momentum-resolved energy-integrated dynamical spin structure factors obtained with unpolarized and polarized neutron scattering~\cite{gaudet2019quantum, smith2022case} are in excellent qualitative agreement with theoretical predictions for $\pi$-flux QSI~\cite{desrochers2023spectroscopic, Desrochers2024Finite, Hosoi2022Uncovering, chern2024pseudofermion}. High-resolution backscattering neutron spectroscopy~\cite{poree2023fractional} results on Ce$_2$Sn$_2$O$_7$ have further highlighted the presence of multiple peaks of decreasing intensity that was recently proposed as a characteristic signature of spinons excitations in $\pi$-flux QSI~\cite{desrochers2023spectroscopic,desrochers2023symmetry, Desrochers2024Finite}. 
\begin{figure}
    \centering
    \includegraphics[width=\linewidth]{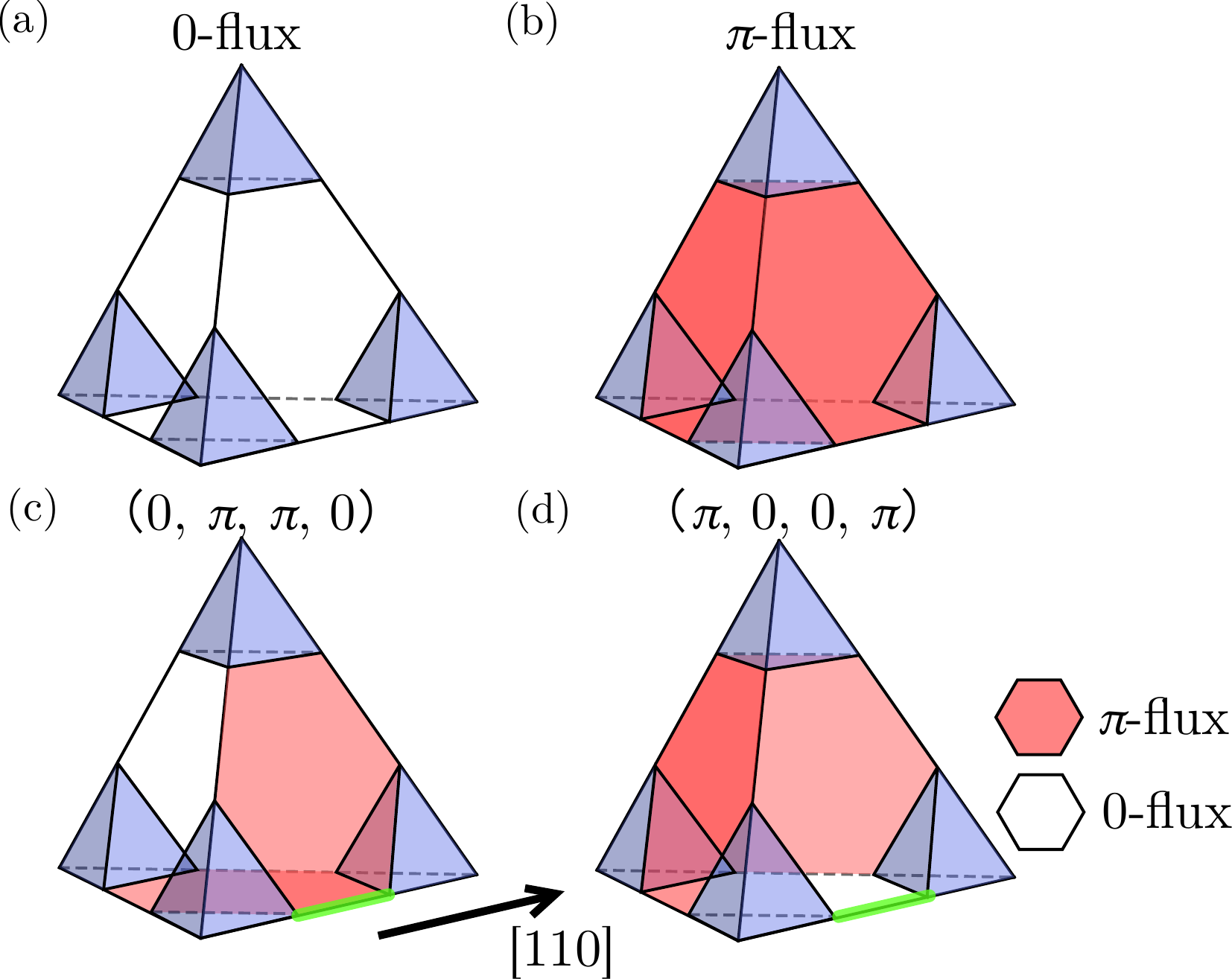}
    \caption{ Hexagonal plaquette fluxes for all allowed phases. (a) 0-flux and (b) $\pi$-flux phase with all plaquettes being threaded by 0 and $\pi$ flux, respectively. (c) $(0,\pi,\pi,0)$ phase where the plaquettes touching the edge highlighted in green are $\pi$-flux and the rests, 0-flux. (d) $(\pi,0,0,\pi)$ phase where the plaquettes touching the edge highlighted in green are 0-flux and the rests, $\pi$-flux.}
    \label{fig:synopsis}
\end{figure}
\begin{figure*}
    \centering
    \includegraphics[width=\linewidth]{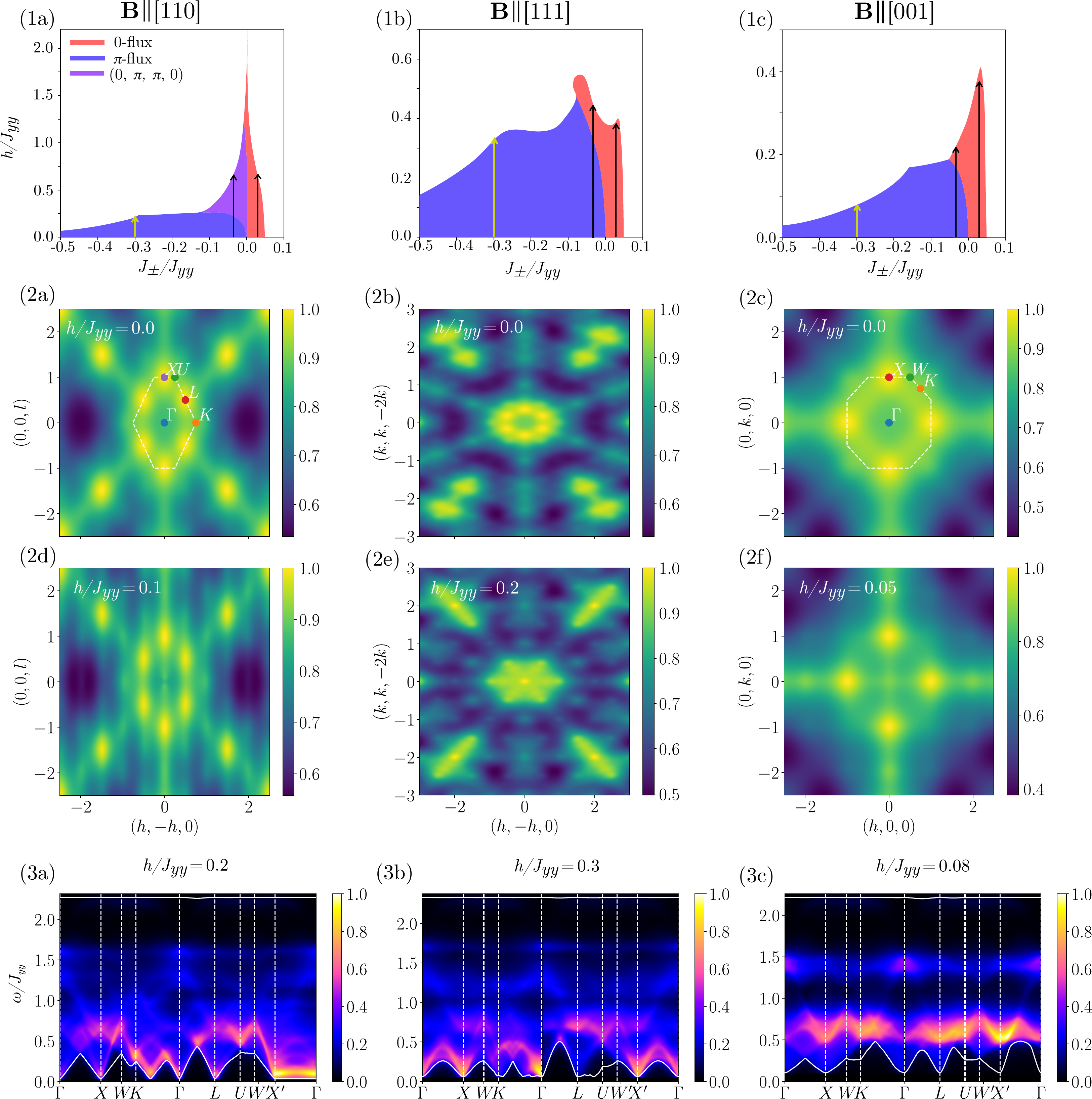}
    \caption{(1) Phase diagram as a function of transverse coupling $J_\pm$ and magnetic field strength $h$ for fields along the (1a) [110], (1b) [111], and (1c) [001] directions. Blue denotes the $\pi$-flux phase, red the $0$-flux phase, and purple the ($0,\pi,\pi,0$) phase. The black arrows at $J_\pm=-0.03$ and $J_\pm=0.03$ and the yellow arrows at $J_\pm=-0.3$ indicate the regions in parameter space where the SSSF and DSSF are calculated. (2) Static spin structure factors in the global frame of $\pi$-flux QSI at $J_\pm=-0.3$ for a [110] field (2a),(2d), a [111] field (2b), (2e), and a [001] field (2c), (2f). (3) Dynamical spin structure factors at finite magnetic fields for a [110] field (3a), a [111] field (3b), and a [001] field (3c).
    }
    \label{fig:synopsis2}
\end{figure*}
Despite these remarkably encouraging results, the unambiguous identification of a QSL is a complex and subtle task that requires the identification of multiple characteristic signatures. 

In this work, we explore more distinctive signatures of DO-QSI that could provide further evidence for the realization of QSI in Ce-based compounds and help guide future experimental efforts. In particular, using the projective symmetry group (PSG) analysis~\cite{wen2002quantum,liu2019competing,liu2021symmetric,schneider2022projective,chern2021theoretical, chern2017fermionic,chern2017quantum, desrochers2023symmetry} and gauge mean field theory (GMFT)~\cite{savary2012coulombic, lee2012generic, savary2016quantum, savary2017disorder, savary2021quantum} on a relevant spin model, we study characteristic response functions in the presence of a magnetic field and their dependence on the magnetic field directions. This is based on the fact that the DO pseudospins couple non-trivially to an external magnetic field~\cite{rau2019frustrated, yao2020pyrochlore,li2017symmetry}. Namely, two pseudospin components ($\tau^x$ and $\tau^y$) have an underlying octupolar magnetic charge density, and $\tau^z$ has a dipolar density such that only $\tau^z$ couples linearly to the magnetic field~\cite{huang2014quantum,yang2020magnetic, rule2006fieldinduced,tian2016fieldinduced,lhotel2017fieldinduced}. The magnetic field orientation should provide a useful tuning knob since the system couples very differently for distinct field orientations. For instance, a field along the [110] direction couples only to one-dimensional spin chains (i.e., the so-called $\alpha$ chains), whereas other perpendicular chains ($\beta$ chains) remain completely decoupled to the magnetic field (see Fig.~\ref{fig:pyrochlore}(e)). In contrast, a field along the [001] direction shown in Fig.~\ref{fig:pyrochlore}(d) couples uniformly in strength to every spin in the system, whereas a [111]-field couples strongly (weakly) to sites forming stacked triangular (kagome) lattices as illustrated in Fig.~\ref{fig:pyrochlore}(f). This fragmentation of the pyrochlore lattice lowers the space group (SG) symmetry and can produce unique observable signatures that may serve as unambiguous evidence for DO-QSI. The application of an external field further offers the thrilling possibility of stabilizing novel exotic QSL for different field orientations. Such a possibility was recently substantiated in Ref.~\cite{yan2023experimentally}, where unconventional phases with staggered and continuously tunable fluxes were argued to be stabilized through perturbative arguments. 

In the rest of the introduction, we quickly discuss some of our main results that are particularly relevant to future neutron scattering experiments. The main text presents more detailed results in different parts of the phase diagram and discussions of the theoretical formulation.

As explained later in detail in Sec.~\ref{sec:DO_ICE}, the pseudospin Hamiltonian of the DO system at zero magnetic fields can be written in terms of new pseudospin variables, namely $S^y \sim \tau^y$ and $S^x, S^z$ that are linear combinations of $\tau^x, \tau^z$. $S^z$ ($S^y$ and $S^x$) transforms like dipolar (octupolar) operators under lattice symmetry transformations. We consider a simplified model:
\begin{equation}
    \mathcal{H_{\text{DO}}} = \sum_{\langle \mathbf{R}_\mu, \mathbf{R}_\nu \rangle} J_{\parallel} S_{\mathbf{R}_\mu}^\alpha S_{\mathbf{R}_\nu}^\alpha - J_\pm (S^+_{\mathbf{R}_\mu} S_{\mathbf{R}_\nu}^- + \text{h.c.}),
\end{equation}
where $J_\parallel=J_{\alpha\alpha}>J_{\beta\beta},\: J_{\gamma\gamma}$ represents the dominant pseudospin exchange parameter of the most general symmetry-allowed XYZ model for the corresponding pseudospin-1/2 component $S^\alpha$, $\alpha=x,\:y,\:\text{or }z$. Here $J_\pm = -(J_{\beta\beta}+J_{\gamma\gamma})/4$ and $\{J_{\alpha\alpha}, J_{\beta\beta}, J_{\gamma\gamma}\}$ is some cyclic permutation of the Ising exchange interactions $\{J_{xx}, J_{yy}, J_{zz}\}$ depending on which is the dominant exchange parameter. As such,  $S^{\pm} = S^\beta \pm i S^\gamma$ if the dominant component is $S^{\alpha}$. A detailed discussion of the above Hamiltonian is featured in Sec.~\ref{sec:DO_ICE}. 
In the GMFT, the pseudospin variables are mapped to an emergent electric field $E$ and vector potential $A$ via $S^\alpha \sim E$ and $S^{\pm} \sim \Phi^{\dagger} e^{\pm i A} \Phi
$, where $\Phi$ represent the bosonic spinons. Furthermore, the emergent photons are described by a $U(1)$ gauge theory~\cite{benton2012seeing, hermele2004pyrochlore, banerjee2008unusual, huang2018dynamics, pace2021emergent, Morampudi2020Spectroscopy}:
\begin{equation}
    \mathcal{H_{\text{eff}}} = U \sum_{\rm links} E^2 +  \sum_{\hexagon} K_{\hexagon} \cos(\nabla \cross A)_{\hexagon}.
    \label{eq:H_QED}
\end{equation}
$\nabla \cross A$ represents the flux of the emergent gauge field $A$ through a hexagonal plaquette. In the absence of magnetic fields, it was shown that in symmetric states respecting all pyrochlore lattice space group symmetries, $\nabla \cross A$ is either zero or $\pi$ in all the hexagonal plaquettes~\cite{desrochers2023symmetry, desrochers2023spectroscopic}, leading to only the 0-flux and $\pi$-flux quantum spin ice states as possible fully symmetric QSL ground states. 

In the presence of magnetic fields, symmetric spin liquids can have more general flux configurations.
With lattice translations and inversion symmetry, which are not broken by the magnetic field along cubic directions, only four hexagonal plaquettes are allowed to have different fluxes (see detailed derivation in Sec.~\ref{sec:magenabled}). As such, we can describe a phase by the fluxes threaded through these four inequivalent hexagonal plaquettes. Such flux configurations can be visualized using a tetrahedron superlattice like the one shown in Fig.~\ref{fig:synopsis}, which captures all four inequivalent plaquettes. Beyond the aforementioned 0- and $\pi$-flux phase in Fig.~\ref{fig:synopsis}(a) and (b), we find that two new exotic phases are allowed only under a [110] field. The phase where the $\pi$-flux plaquettes have normal vectors perpendicular to [110] is referred to as $(0,\pi,\pi,0)$ (see Fig.~\ref{fig:synopsis}(c)). On the other hand, the phase where the $0$-flux plaquettes have normal vectors perpendicular to [110] is referred to as $(\pi,0,0,\pi)$ in Fig.~\ref{fig:synopsis}(d).

\begin{figure*}
    \centering
    \includegraphics[width=\linewidth]{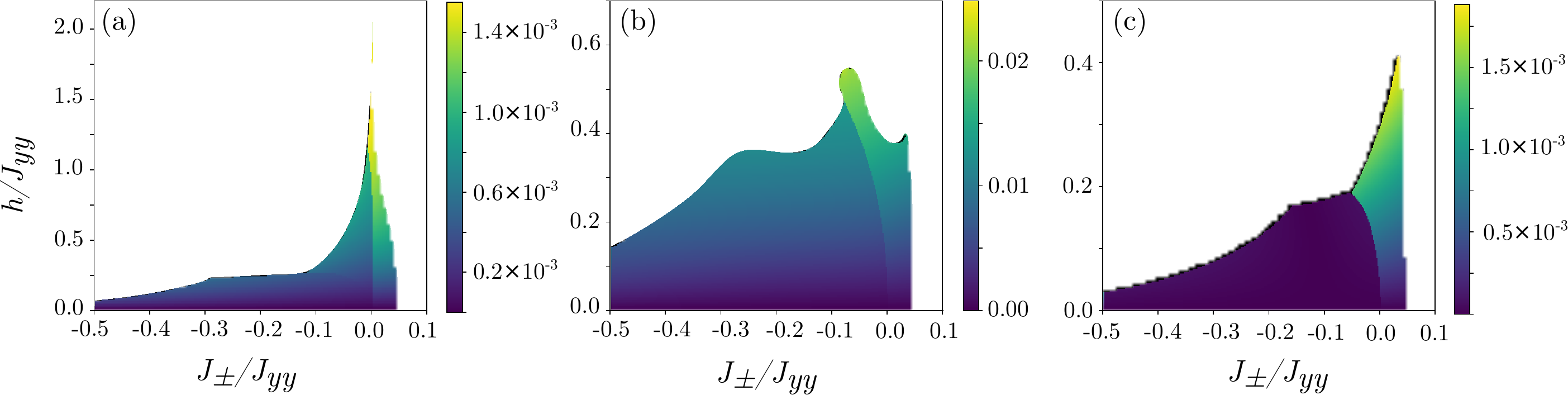}
    \caption{Normalized magnetization $g_{zz}\mu_B$ per pyrochlore site along the applied magnetic field, $|\mathbf{m}|=\sum_\mu \hat{\mathbf{n}}\cdot \hat{\mathbf{z}}_\mu\langle S^z_\mu \rangle/4$, with $\mathbf{n}$ parallel to (a) [110], (b) [111], and (c) [001].}
    \label{fig:magnetizations}
\end{figure*}

We construct the phase diagrams for magnetic fields along [110], [111], and [001] directions as shown in Fig.~\ref{fig:synopsis2}. Under a non-zero [110] field, the $(0,\pi,\pi,0)$ state is shown to be stable near the Ising point ($J_\pm=0$) up to a magnetic field of $h/J_{yy}\approx 1$ in Fig.~\ref{fig:synopsis2}(1a). We observe a first-order field-induced phase transition from the $\pi$-flux phase to the $(0,\pi,\pi,0)$ phase close to the Ising point as evidenced by the abrupt jump in magnetization $|\mathbf{m}|$ shown in Fig.~\ref{fig:magnetizations}(a) and Fig.~\ref{fig:magn}. In the same light, as shown in Fig.~\ref{fig:synopsis2}(1b) and (1c), fields along the [111] and [001] can again induce a phase transition near the Ising point, this time into the 0-flux phase. This transition is also first-order, as seen in Fig.~\ref{fig:magnetizations}(b) and (c).

However, the exciting prospect of seeing exotic QSL to QSL field-induced transitions, as well as novel staggered flux phases, have to be carefully weighed against realistic expectations from current material candidates. For DO-QSI candidate Ce$_2$Zr$_2$O$_7$, the estimated microscopic exchange parameters fall near $J_\pm \approx 0.3 J_{\parallel}$~\cite{smith2023dipole,smith2023quantum,bhardwaj2022sleuthing}, where $J_{\parallel}=J_{yy}$ or $J_{\parallel}=J_{xx}$, which leads to the $\pi$-flux QSI at zero magnetic field. The critical magnetic fields are therefore quite small: using the estimated $g$-factor, $g=2.24$, in Ref.~\cite{smith2023quantum}, $h_c\approx 0.11$~T for $[110]$ direction; $h_c\approx 0.17$~T for $[111]$ direction; $h_c\approx 0.05$~T for $[001]$ direction. Furthermore, for such large transverse couplings, one cannot see any transition to the 0-flux or $(0,\pi,\pi,0)$-flux phases. Therefore, such a transition is not directly relevant to Ce$_2$(Zr,Sn,Hf)$_2$O$_7$, but it remains a remarkable observation that could be pertinent to future candidate materials. 

To further substantiate the experimental relevance of our findings, we computed the static spin structure factor (SSSF) and the dynamic spin structure factor (DSSF) using the estimated microscopic exchange parameters for Ce$_2$Zr$_2$O$_7$. These results provide direct comparisons with neutron scattering experiments as reasoned in Sec.~\ref{sec:experimentalsig}. As such, we construct a distinguishing set of experimental signatures for $\pi$-flux QSI under various field orientations directly applicable to Ce$_2$Zr$_2$O$_7$. We summarize these experimentally relevant theoretical predictions below.

For the SSSF, we find that, under a [110] field, stripe patterns emerge, as shown in Fig.~\ref{fig:synopsis2}(2d). This is due to the fragmentation of the pyrochlore lattice into the one-dimensional $\alpha$ and $\beta$ chains. Furthermore, we also observe heightened intensities along $(0,0,l)$ with an emergent ``pinch point" at the $\Gamma$ point. As discussed in Sec.~\ref{sec:h110}, these signatures originate from the polarization of the strongly coupled $\alpha$ chains. Under a [111] field, the DO-QSI exhibits Kagome-ice-like correlations illustrated in Fig.~\ref{fig:synopsis2}(2e). Finally, under a [001] field, since the polarized paramagnetic configuration is still 2-in-2-out in the $|S^z=\pm\rangle$ basis, the rod-like motifs, which are associated with $\pi$-flux DO-QSI~\cite{desrochers2023spectroscopic, castelnovo2019rod}, persist as shown by the cross-like signature in Fig.~\ref{fig:synopsis}(2f).

Previous studies~\cite{desrochers2023spectroscopic,desrochers2023symmetry} have shown that the DSSF of $\pi$-flux QSI has three distinctive peaks originating from the two flat spinon bands (each of them is doubly degenerate) at zero magnetic fields. 
It turns out that, under any finite magnetic field, the double degeneracy of the spinon bands is lifted while the flatness of the dispersion remains mostly intact. This leads to the splitting of the three-peak structure into multiple peaks in the DSSF, as explained in Sec.~\ref{sec:DSSF}. For Ce$_2$Zr$_2$O$_7$ where the transverse coupling $J_\pm$ is large, we show that under the [111] and [110] fields, as shown in Fig.~\ref{fig:synopsis2}(3a)-(3b), the three peaks splits into five, whereas under a [001] field we see virtually no splitting at all as seen in Fig.~\ref{fig:synopsis2}(3c). These features are more clearly captured in Fig.~\ref{fig:2SpinonDOS111Jpm-0.03}(d)-(f) as noted in Sec.~\ref{sec:DSSF}. Furthermore, the size of the splitting of multiple peaks strongly depends on the field directions. This strong dependence makes up a unique experimental profile of $\pi$-flux QSI. However, it is worth noting that this absolute energy scale of the splitting is quite small given that $J_{yy} \approx 0.06$~meV for Ce$_2$Zr$_2$O$_7$. 
As such, experimental resolutions of these fine features may be challenging.

The structure of the following sections is as follows. 
In Sec.~\ref{sec: Formalism}, we explain the underlying microscopic degrees of freedom in DO systems and introduce GMFT formalism. In Sec.~\ref{sec:PSGClass}, we perform the PSG classification for possible quantum spin ice states for different field directions and compute the corresponding phase diagrams in Sec.~\ref{sec:phase_diagram}. In Sec.~\ref{sec:SSSF} and \ref{sec:DSSF}, we present the results of the static and dynamical spin structure factors in various regimes in the phase diagram before concluding in Sec.~\ref{sec:discussion}.

\section{\label{sec: Formalism} Formalism}

\subsection{Dipolar-Octupolar Spin Ice\label{sec:DO_ICE}}

The magnetic properties of the compounds A$_2$B$_2$O$_7$ depend on the rare-earth ion A$^{3+}$~\cite{gardner2010magnetic,greendan2006rare} that form a pyrochlore lattice illustrated in Fig.~\ref{fig:pyrochlore}(a). The single-ion electronic states are split by spin-orbit coupling and a $D_{3d}$-symmetric crystalline electric field. For magnetic ions with an odd number of electrons, such as Ce$^{3+}$, all states must be at least twofold degenerate by time-reversal symmetry. This splitting may result in a well-isolated ground state doublet. In a crystalline environment with a $D_{3d}$ symmetry, there are two possibilities for these low-lying doublets: (i) the effective spin-$1/2$ case, where the doublet transforms as the $\Gamma^4$ irreducible representation, and (ii) a dipolar-octupolar (DO) doublet that is built from the one-dimensional irreducible representations $\Gamma^5\oplus\Gamma^6$ that do not mix under spatial transformations~\cite{rau2019frustrated, huang2014quantum}. Ce$_2$(Zr,Sn,Hf)$_2$O$_7$ all have a well-isolated doublet that is dipolar-octupolar. This ground state doublet determines the low-energy physics and can be conveniently represented by the pseudospins-$1/2$ operators
\begin{subequations}\label{eq:gamma5gamma6}
\begin{align}
&\tau^x=\mathcal{C}_0\mathcal{P}\left(\left(J^x\right)^3-\overline{J^x J^y J^y}\right) \mathcal{P} \\
& \tau^y=\mathcal{C}_1 \mathcal{P}\left(\left(J^y\right)^3-\overline{J^y J^x J^x}\right) \mathcal{P} \\
& \tau^z=\mathcal{C}_2 \mathcal{P} J^z \mathcal{P},
\end{align}
\end{subequations}
where $\mathcal{C}_0$, $\mathcal{C}_1$, $\mathcal{C}_2$ are CEF parameters, $J^a$ are angular momentum components, the overline indicates symmetrized products, and $\mathcal{P}$ is the projection down to the ground state doublet. The different components are defined in a sublattice-dependent local frame (see Appendix~\ref{sec:Appendix_DOice}) where the local $z$-axis point out of the up-pointing tetrahedrons as illustrated in Fig.~\ref{fig:pyrochlore}(b). The underlying magnetic charge density of the $\tau^x$ and $\tau^y$ are octupolar, whereas $\tau^z$ is dipolar. 

Let us then construct the most general Hamiltonian coupled to an external magnetic field by first examining the pseudospin transformation under symmetry operations of the pyrochlore lattice. The generators of the pyrochlore space group are lattice translations by the basis vectors $\mathbf{e}_1$, $\mathbf{e}_1$ and $\mathbf{e}_3$ ($T_1$, $T_1$, and $T_3$), a rotoinversion along the [111] axis ($\bar{C}_6$), and a screw operation along the [110] direction ($S$)~\cite{hahn1983international}. The pseudospin components transform under these generators as:
\begin{subequations}\label{eq:symopXYZ}
\begin{align}
T_i: \{\tau^x_{\mathbf{R}_\mu}, \tau^y_{\mathbf{R}_\mu}, \tau^z_{\mathbf{R}_\mu}\} &\rightarrow  \{\tau^x_{T_i(\mathbf{R}_\mu)}, \tau^y_{T_i(\mathbf{R}_\mu)}, \tau^z_{T_i(\mathbf{R}_\mu)}\}\\
\bar{C}_6: \{\tau^x_{\mathbf{R}_\mu}, \tau^y_{\mathbf{R}_\mu}, \tau^z_{\mathbf{R}_\mu}\} &\rightarrow  \{\tau^x_{\bar{C}_6(\mathbf{R}_\mu)}, \tau^y_{\bar{C}_6(\mathbf{R}_\mu)}, \tau^z_{\bar{C}_6(\mathbf{R}_\mu)}\}\\
S: \{\tau^x_{\mathbf{R}_\mu}, \tau^y_{\mathbf{R}_\mu}, \tau^z_{\mathbf{R}_\mu}\} &\rightarrow  \{-\tau^x_{S(\mathbf{R}_\mu)}, \tau^y_{S(\mathbf{R}_\mu)}, -\tau^z_{S(\mathbf{R}_\mu)}\},
\end{align}
\end{subequations}
where $\mathbf{R}_{\mu}$ labels the pyrochlore sites with index $\mu\in\{0,1,2,3\}$ identifying the sublattice as shown in Fig.~\ref{fig:pyrochlore}(b). We see that $\tau^x$ and $\tau^z$ transform as dipoles, whereas $\tau^y$ transforms as an octupole. For this reason, $\tau^x$ and $\tau^z$ are commonly referred to as dipolar pseudospin components even though $\tau^x$ has an octupolar magnetic charge density (see Eq.~\eqref{eq:gamma5gamma6}). The most general symmetry-allowed Hamiltonian with nearest neighbor coupling and an external magnetic field is then
\begin{align}
    \mathcal{H} =& \sum_{\langle \mathbf{R}_\mu,\mathbf{R}_\nu \rangle} \left[J_{xx} \tau_{\mathbf{R}_\mu}^x \tau_{\mathbf{R}_\nu}^x +J_{yy} \tau_{\mathbf{R}_\mu}^y \tau_{\mathbf{R}_\nu}^y + J_{zz} \tau_{\mathbf{R}_\mu}^z \tau_{\mathbf{R}_\nu}^z \right.\nonumber\\
    &\left.\quad\quad\quad +J_{xz}\left(\tau_{\mathbf{R}_\mu}^x \tau_{\mathbf{R}_\nu}^z + \tau_{\mathbf{R}_\mu}^z \tau_{\mathbf{R}_\nu}^x\right) \right]\nonumber\\
    &- \mu_B \sum_{\mathbf{R}_\mu} \left[ (\mathbf{B} \cdot \hat{\mathbf{z}}_{\mu}) \left(g_{xx}\tau^x_{\mathbf{R}_\mu} + g_{zz}  \tau^z_{\mathbf{R}_\mu} \right) \right. \nonumber\\
    &\left. \quad\quad\quad\quad + g_{yy} ((B^y_{\mu})^3-3(B^x_{\mu})^2B^y_\mu)\tau^y_{\mathbf{R}_\mu} \right],
\end{align}
where $\hat{\mathbf{z}}_{\mu}$ are the local pseudospin $z$-axis on sublattice $\mu$ (see Fig.~\ref{fig:pyrochlore}(b) and table \ref{tab: Local basis}) and $B^{a}_\mu$ refers to the $a$-component of the magnetic field as defined in the local frame of the $\mu^{\text{th}}$ sublattice. In the following, we will consider the weak-field limit where we can drop the cubic magnetic field coupling to $\tau^{y}$. Because the underlying magnetic charge density of $\tau^x$ is octupolar, as shown in equation~\eqref{eq:gamma5gamma6}, $g_{xx}\approx 0$ and only $g_{zz}$ is non-zero. The mixing term $J_{xz}$ can further be removed through a rotation about the local $y$-axis to bring the Hamiltonian into a simple XYZ form. Such a rotation takes the form $\tau^y=S^y$, $\tau^x=\cos (\theta) S^x-\sin (\theta) S^z$ and $\tau^z=\sin (\theta) S^x+\cos (\theta) S^z$, where $\theta$ depends on the exchange coupling constants. However, since experimental studies showed that $J_{xz}$ is small in the leading candidate materials Ce$_2$Zr$_2$O$_7$~\cite{smith2022case}, we will assume $J_{xz}=0$ for the purpose of this study (i.e., $\theta=0$). Furthermore, in order to clearly discern the effect of magnetic fields, we consider a simpler XXZ case where $J_{yy}>J_{xx}=J_{zz}$ and $J_{yy}>0$. The choice of $J_{yy}$ as the leading coupling constant is also motivated by results on Cerium compounds.  It is worth mentioning that recent studies also propose that $J_{xx}$ could be the dominant interaction with $J_{xz}\ne 0$~\cite{gao2024emergent}. However, we leave such a case for future studies and discuss it further in Sec.~\ref{sec:discussion}. After all these considerations, the Hamiltonian we will be interested in is 
\begin{equation}
\begin{aligned}
    \mathcal{H_{\text{XXZ}}} &= \sum_{\langle \mathbf{R}_\mu, \mathbf{R}_\nu \rangle} J_{yy} S_{\mathbf{R}_\mu}^y S_{\mathbf{R}_\nu}^y - J_\pm (S^+_{\mathbf{R}_\mu} S_{\mathbf{R}_\nu}^- + \text{h.c.}) \\
    &\qquad - \sum_{\mathbf{R}_\mu} h(\hat{\mathbf{n}} \cdot \hat{\mathbf{z}}_{\mu}) S^z_{\mathbf{R}_\mu}\label{eq:H_XXZ},
\end{aligned}
\end{equation}
where $J_\pm = -(J_{xx}+J_{zz})/4$, $h=g_{zz}\mu_B |\mathbf{B}|$, and $\hat{\mathbf{n}}$ is the direction of the magnetic field. 

In the Ising limit (i.e., $J_{\pm}/J_{yy}\to 0$ and $h/J_{yy}\to 0$), the above antiferromagnetic coupling $J_{yy}$ energetically enforces the sum over the $y$-component of the pseudospin to be zero for every tetrahedron. This set of local constraints, commonly referred to as the ice rules, leads to a classical spin liquid with an extensive ground state degeneracy. The addition of a small transverse term $J_{\pm}$ can be treated perturbatively, upon which the system has been shown to stabilize 0-flux QSI for $J_{\pm}>0$ and $\pi$-flux QSI for $J_{\pm}<0$~\cite{hermele2004pyrochlore, banerjee2008unusual, shannon2012quantum, kato2015numerical}.

\subsection{Gauge mean-field Theory\label{sec:GMFT}}

Gauge mean-field theory (GMFT) is a slave particle formalism that describes QSI beyond the perturbative Ising regime by expressing $\mathcal{H}_{\text{XXZ}}$ as a $U(1)$ compact gauge theory with matter fields, where spinons can be directly studied~\cite{savary2012coulombic, lee2012generic, savary2016quantumspinliquids, li2017symmetry, yao2020pyrochlore, desrochers2023symmetry}.

The construction goes as follows: First, we introduce a slave matter field $Q\in \mathbb{Z}$ on the parent diamond lattice, as depicted in Fig.~\ref{fig:pyrochlore}(a) by the spheres at centers of the tetrahedrons. Here, we label the parent diamond lattice sites with the sublattice indexed diamond coordinates (SIDC), $\mathbf{r}_\alpha$, defined in equation~\eqref{eq:SIPC_SIDC}, where $\alpha= A, B$ denotes the diamond sublattice index (A refers to the centers of down-pointing tetrahedrons, and B, that of up-pointing tetrahedrons), as shown in Fig.~\ref{fig:pyrochlore}(b). The charges $Q$ conceptually correspond to tetrahedrons that break the ice rules described above since they have to respect the following Gauss' law 
\begin{equation}
    Q_{\mathbf{r}_\alpha} = \sum_{i\in \partial t_{\mathbf{r}_\alpha}} S_{\mathbf{R}_\mu}^y,\label{eq:gaussrule}
\end{equation}
where $t_{\mathbf{r}_\alpha}$ denotes the tetrahedron centered at $\mathbf{r}_\alpha$ and $\partial t_{\mathbf{r}_\alpha}$ refers to the four pyrochlore sites forming its boundary. The associated charge raising and lowering operators can then be naturally defined using the conjugate variable $\varphi_{\mathbf{r}_\alpha}$ (i.e., $[\varphi_{\mathbf{r}_\alpha}, Q_{\mathbf{r}_\alpha'}]=i\delta_{\mathbf{r}_\alpha \mathbf{r}_\alpha'}$) to be $\Phi_{\mathbf{r}_\alpha}^\dagger = e^{i\varphi_{\mathbf{r}_\alpha}}$ and $\Phi_{\mathbf{r}_\alpha} = e^{-i\varphi_{\mathbf{r}_\alpha}}$. By construction, the length of this operator is $|\Phi_{\mathbf{r}_\alpha}^\dagger \Phi_{\mathbf{r}_\alpha}| = 1$. 

Next, the original pseudospin operators can be extended to act on the enlarged Hilbert space as follows:
\begin{equation} \label{eq:mapping_pseudospin_eQED}
\begin{aligned}
    &S_{\mathbf{R}_\mu}^{+} \rightarrow \Phi^\dagger_{\mathbf{r}_A} \left( \frac{1}{2}e^{iA_{\mathbf{r}_A, \mathbf{r}_A + \mathbf{b}_\mu}} \right)\Phi_{\mathbf{r}_A+\mathbf{b}_\mu}\\
    &S_{\mathbf{R}_\mu}^{y} \rightarrow E_{\mathbf{r}_A, \mathbf{r}_A+\mathbf{b}_\mu},
\end{aligned}
\end{equation}
where $E_{\mathbf{r}_A, \mathbf{r}_A+\mathbf{b}_\mu}$ and $A_{\mathbf{r}_A, \mathbf{r}_A + \mathbf{b}_\mu}$ are canonical conjugate electric and gauge fields acting on the initial spin Hilbert space, $\mathbf{b}_\mu$ are vectors connecting A diamond sites to the nearest four B diamond sites defined in Eq.~\eqref{eq:bmu}, and $\mathbf{R}_\mu$ is the pyrochlore site connecting the up and down tetrahedron centered at $\mathbf{r}_A$ and $\mathbf{r}_A+\mathbf{b}_{\mu}$ respectively. To recover the initial physical spin Hilbert space, the Gauss' law of Eq.~\eqref{eq:gaussrule} has to be enforced at every tetrahedron.

After this exact mapping, the Hamiltonian becomes
\begin{equation}
\begin{aligned}\label{eq:H_parton}
    \mathcal{H}_{\text{parton}} = &\frac{J_{yy}}{2}\sum_{\mathbf{r}_\alpha} Q_{\mathbf{r}_\alpha}^2 - \frac{J_\pm}{4} \sum_{\mathbf{r}_\alpha, \mu\neq\nu} \Phi^\dagger_{\mathbf{r}_\alpha + \eta_\alpha \mathbf{b}_\mu} \Phi_{\mathbf{r}_\alpha + \eta_\alpha \mathbf{b}_\nu} \\
    & \times e^{i\eta_\alpha (A_{\mathbf{r}_\alpha, \mathbf{r}_\alpha +\eta_\alpha \mathbf{b}_\mu} - A_{\mathbf{r}_\alpha, \mathbf{r}_\alpha +\eta_\alpha \mathbf{b}_\nu})} - \frac{h}{4} \sum_{\mathbf{r}_A, \mu} (\hat{\mathbf{n}} \cdot \hat{\mathbf{z}}_{\mu}) \\&\times\left(\Phi^\dagger_{\mathbf{r}_A} \Phi_{\mathbf{r}_A + \mathbf{b}_\mu} e^{iA_{\mathbf{r}_A, \mathbf{r}_A+\mathbf{b}_\mu}} + h.c.\right),
\end{aligned}
\end{equation}
which has and a $U(1)$ gauge structure with the following gauge transformation
\begin{equation}
\begin{aligned}
&\Phi_{\mathbf{r}_\alpha} \rightarrow \Phi_{\mathbf{r}_\alpha} e^{i \chi_{\mathbf{r}_\alpha}} \\
&A_{\mathbf{r}_\alpha, \mathbf{r}_\beta^{\prime}} \rightarrow A_{\mathbf{r}_\alpha, \mathbf{r}_\beta^{\prime}}-\chi_{\mathbf{r}_\beta^{\prime}}+\chi_{\mathbf{r}_\alpha}.\label{eq:gaugetransformation}
\end{aligned}
\end{equation}
With the above mapping, the initial spin Hamiltonian is now described by spinons coupled to an emergent compact $U(1)$ gauge field $A$. The $J_{\pm}$ term describes spinons hopping within the same sublattices (i.e., intra-sublattice hopping), and the Zeeman term enables inter-sublattice hopping. We see that this construction does not rely on any perturbative arguments and allows us to capture physics away from the 2-in-2-out spin ice manifold where defect tetrahedra are important and need to be considered.

To obtain a more tractable model, we make two approximations. 
(1) A saddle point approximation where the gauge field $A_{\mathbf{r}_A, \mathbf{r}_A+\mathbf{b}_\mu}$ is fixed to a constant background $\overline{A}_{\mathbf{r}_A, \mathbf{r}_A+\mathbf{b}_\mu}$. After this strong approximation, the matter field $Q$ and the gauge field are decoupled. Accordingly, the lattice Gauss's law in equation~\eqref{eq:gaussrule} needs not to be respected, and the $Q$ field can be integrated out. (2) A large-$N$ approximation where the operator identity $|\Phi^\dagger_{\mathbf{r}_\alpha} \Phi_{\mathbf{r}_\alpha}|=1$ is relaxed to $\langle \Phi^\dagger_{\mathbf{r}_\alpha} \Phi_{\mathbf{r}_\alpha} \rangle = \kappa$. This constraint can be enforced by a sublattice-dependent Lagrange multiplier $\lambda_{\alpha}$. One important note here is that the large-$N$ approximation with the naive choice $\kappa=1$ predicts a critical $J_{\pm}^c\approx 0.2 J_{yy}$ for $0$-flux QSI to transition into the all-in-all-out (AIAO) ordered state, whereas sign-problem-free Quantum Monte-Carlo (QMC) studies find that $J_{\pm}^c \approx 0.05 J_{yy}$. This discrepancy is an artifact of the large-$N$ approximation and can be remedied by choosing $\kappa=2$. For those reasons, we use $\kappa=2$ throughout the rest of this article. This choice was further shown to cure many important physical inconsistencies present in the $\kappa=1$ theory~\cite{desrochers2023symmetry, desrochers2023spectroscopic}. After such approximations, we finally arrive at an exactly solvable Hamiltonian. The last remaining ambiguity is how to fix the gauge field background $\bar{A}$. We show in the following section how the gauge field background can be fixed by enumerating all configurations that respect a given set of symmetries through a projective symmetry group analysis. 
\section{PSG Classification with different field directions \label{sec:PSGClass}}
\subsection{Generalities}

Due to the emergent gauge structures~\eqref{eq:gaugetransformation}, which originates from the Gauss law constraint~\eqref{eq:gaussrule},  mean-field Ans\"atze needs only to respect the lattice symmetries up to a gauge transformation. This is because mean-field Ans\"atze related via a gauge transformation correspond to the same physical states.

To see this, first consider a gauge transformation of the form~\eqref{eq:gaugetransformation}, which is generated by the operator
\begin{equation}
    U(\{\chi\})=\prod_{\mathbf{r}_\alpha} \exp \left(i \chi_{\mathbf{r}_\alpha}\left(Q_{\mathbf{r}_\alpha}-\sum_\mu E_{\mathbf{r}_\alpha, \mathbf{r}_\alpha+\eta_\alpha \mathbf{b}_\mu}\right)\right).
\end{equation}
Such a transformation acts trivially on any physical state that respects the Gauss' law~\eqref{eq:gaussrule}. After the saddle point approximation, the gauge structure is lost and a given MF gauge field configuration $\{\bar{A}_{\mathbf{r}_\alpha, \mathbf{r}_\beta'}\}$ now transforms as $U:\{\bar{A}_{\mathbf{r}_\alpha, \mathbf{r}_\beta'}\}\rightarrow\{\bar{A}_{\mathbf{r}_\alpha, \mathbf{r}_\beta'} +\eta_{\alpha}(\chi_{\mathbf{r}_\beta^{\prime}}-\chi_{\mathbf{r}_\alpha})\}$. To get a physical spin wave function back, we must project down to states that respect the lattice Gauss's law using some projection operator $\mathcal{P}_\text{gauss}$. Since $U$ acts trivially on the projected (physical) state, the gauge transformation and projection operation commute: $[U, \mathcal{P}_{\text{gauss}}]=0$. Therefore, we can see that 
\begin{equation}
    \mathcal{P}_\text{gauss} U|\Psi\rangle = U\mathcal{P}_\text{gauss}|\Psi\rangle = \mathcal{P}_\text{gauss}|\Psi\rangle, 
\end{equation} 
or that any two wavefunctions $|\Psi\rangle$ and $U|\Psi\rangle$ connected through a gauge transformation $U$ results in the same physical state. 

The above implies that when considering a symmetry operation $\mathcal{O}$ acting on the Hamiltonian 
\begin{equation}
\begin{aligned}
& \mathcal{H}_{\mathrm{GMFT}}\left(\left\{\bar{A}_{\mathbf{r}_\alpha,\mathbf{r}_\beta'}\right\}\right)|\Psi_{\mathbf{r}_\alpha}\rangle \\
&\quad\rightarrow \mathcal{O} \mathcal{H}_{\mathrm{GMFT}}(\{\bar{A}\}) \mathcal{O}^{\dagger} \mathcal{O} |\Psi_{\mathbf{r}_\alpha}\rangle \\
& \quad\quad = \mathcal{H}_{\mathrm{GMFT}}(\{\bar{A}_{\mathcal{O}(\mathbf{r}_\alpha),\mathcal{O}(\mathbf{r}'_\beta)}\})|\Psi_{\mathcal{O} (\mathbf{r}_\alpha)}\rangle ,
\end{aligned}
\end{equation}
$\mathcal{O}$ still maps the Ans\"atz to the same physical state if there exists a gauge transformation $U$ such that $U: \mathcal{H}_{\mathrm{GMFT}}(\{\bar{A}_{\mathcal{O}(\mathbf{r}_\alpha),\mathcal{O}(\mathbf{r}'_\beta)}\})\rightarrow  \mathcal{H}_{\mathrm{GMFT}}(\{\bar{A}_{\mathbf{r}_\alpha,\mathbf{r}'_\beta}\})$. The PSG is then enlisted as a tool to list all inequivalent mean-field Ans\"atze that respect a given set of symmetries. In particular, it can enumerate all fully symmetric QSLs that respect all lattice symmetries. Notice that the PSG solution is dependent on the symmetries imposed. This implies that in the presence of a magnetic field, new fully symmetric states may be allowed if the field breaks some spatial symmetries. 

In the absence of a magnetic field, previous analyses have shown that the only two fully symmetric Ans\"atze of the XXZ model for DO-QSI are the 0- and $\pi$-flux phases~\cite{ desrochers2023spectroscopic}. Below, we examine whether new fully symmetric phases are allowed in the presence of a magnetic field.

\subsection{PSG Classification}

\subsubsection{Flux Description}\label{sec:fluxdescription}
For the purpose of this discussion, it is important to be precise about the use of lattice curl of the gauge field, which we refer to as fluxes of the hexagonal plaquettes. As we see in Fig.~\ref{fig:pyrochlore}(c), a traversal in the hexagon involves three different pyrochlore sublattices. Therefore, we can explicitly write out the lattice curl involving three sublattices with indices $\mu,\nu,\lambda$ as
\begin{equation}
\begin{aligned}
     F_{\mu\nu\lambda} (\mathbf{r}_A)=
     & -\bar{A}_{\mathbf{r}_A+\mathbf{e}_\lambda,\mathbf{r}_A+\mathbf{e}_\lambda+\mathbf{b}_\mu}+\bar{A}_{\mathbf{r}_A+\mathbf{e}_\lambda,\mathbf{r}_A+\mathbf{e}_\lambda+\mathbf{b}_\nu} \\
     &-\bar{A}_{\mathbf{r}_A+\mathbf{e}_\nu,\mathbf{r}_A+\mathbf{e}_\nu+\mathbf{b}_\lambda} + \bar{A}_{\mathbf{r}_A+\mathbf{e}_\nu,\mathbf{r}_A+\mathbf{e}_\nu+\mathbf{b}_\mu}\\
     &-\bar{A}_{\mathbf{r}_A+\mathbf{e}_\lambda,\mathbf{r}_A+\mathbf{e}_\lambda+\mathbf{b}_\nu}+\bar{A}_{\mathbf{r}_A+\mathbf{e}_\lambda,\mathbf{r}_A+\mathbf{e}_\lambda+\mathbf{b}_\lambda},\label{eq:plaquettealgebra}
\end{aligned}
\end{equation}
where $\mathbf{e}_{\mu}$ are lattice basis vectors (see Appendix~\ref{sec:AppendixPyrochlore} for definitions). This is useful for describing the physical states since fluxes are gauge-invariant observables. A physical QSI state of the XXZ model is uniquely defined by a configuration of all hexagonal plaquette fluxes $\{F_{\mu\nu\lambda}(\mathbf{r}_A)\}$. 

\subsubsection{PSG Classification in the absence of a field}
Without any magnetic field, The Hamiltonian is invariant under the pyrochlore SG. The generators of the SG act on the SIDC (see Appendix~\ref{sec:AppendixPyrochlore}) as
\begin{subequations}\label{eq:PyrochloreSGgen}
\begin{align}
T_i&: \mathbf{r}_\alpha \mapsto\left(r_1+\delta_{i, 1}, r_2+\delta_{i, 2}, r_3+\delta_{i, 3}\right)_\alpha \\
\bar{C}_6&: \mathbf{r}_\alpha \mapsto\left(-r_3,-r_1,-r_2\right)_{\pi(\alpha)} \\
S&: \mathbf{r}_\alpha \mapsto\left(-r_1,-r_2, r_1+r_2+r_3+\delta_{\alpha, A}\right)_{\pi(\alpha)}.
\end{align}
\end{subequations}
Since the latter two operations swap the diamond sublattice, we define the permutation operator $\pi(A)=B$, $\pi(B)=A$. As mentioned previously, there are only two fully symmetric $U(1)$ QSLs for the XXZ model of interest that respect all pyrochlore SG operations. These are the $0$- and $\pi$-QSI states where the hexagonal plaquettes are all threaded by a static 0 or $\pi$-flux, respectively. 

On the other hand, the pseudospin components $S^z$ on each sublattice couple differently to the magnetic field (see Eq.~\eqref{eq:H_XXZ}). Transformations that interchange tetrahedron sublattices that couple differently are thus no longer spatial symmetries in the presence of a field. As such, new fully symmetric states may emerge and can be categorized using the PSG analysis.

\subsubsection{Magnetically Enabled Symmetric Spin Liquids \label{sec:magenabled}}

Even in the presence of a field pointing in an arbitrary direction, both translations $T_i$ and inversion symmetries $I=\bar{C}_6^3$ are always present. This is because neither transformations exchange pyrochlore sublattices. Before studying specific field directions, let us consider the constraints imposed by translation and inversion on the fluxes of symmetric Ans\"atze. 

Under lattice translation symmetries, the PSG analysis shows that the plaquettes of all fully symmetric states must be translationally invariant (see Appendix~\ref{sec:AppendixPSG}). The resulting equivalence classes are:
\begin{equation}
\begin{aligned}
[F_{\mu\nu\lambda}(\mathbf{0}_A)]&=\{F_{\mu\nu\lambda} (\mathbf{r}_A)\}=[\bar{F}_{\mu\nu\lambda}]=[(-1)^p\bar{F}_{\sigma(\mu\nu\lambda)}],
\end{aligned}
\end{equation}
where we choose the flux at $\mathbf{0}_A$ to represent all other fluxes related via lattice translations. Here $\sigma$ is any permutation of the three sites, and $p$ is the parity of the permutation. Furthermore, it is found that upon imposing $I$, plaquette fluxes of fully symmetric states can only be 0 or $\pi$ (See details in Appendix~\ref{sec:AppendixPSG0}). This implies $[\bar{F}_{\mu\nu\lambda}]=[\bar{F}_{\sigma(\mu\nu\lambda)}]$, since fluxes are defined modulo $2\pi$. Any symmetric state of the XXZ model that respects translation symmetries and inversion is then specified by four plaquettes fluxes $([\bar{F}_{123}], [\bar{F}_{320}], [\bar{F}_{301}], [\bar{F}_{102}])$ that must be either 0 or $\pi$. There are a total of $2^{4}=16$ different inequivalent states of the XXZ model described by GMFT that are translational- and inversion-invariant. Now, let us consider specific field directions for which additional symmetries might be present. 

First, under a $[111]$ field, the Zeeman couplings on the four sublattices are $-h(\hat{\mathbf{n}}\cdot\hat{\mathbf{z}}_0, \hat{\mathbf{n}}\cdot\hat{\mathbf{z}}_1, \hat{\mathbf{n}}\cdot\hat{\mathbf{z}}_2, \hat{\mathbf{n}}\cdot\hat{\mathbf{z}}_3)=-\frac{h}{3}(3,-1,-1,-1)$. Space group operations that interchange sublattices 1, 2, and 3 are thus symmetry operations, which are generated by $\bar{C}_6$ and $T_i$. PSG analysis reveals that there are two allowed fully symmetric states with flux configuration $(n_1\pi, n_1\pi, n_1\pi, n_1\pi)$, where $n_1\in \{0, 1\}$ (see Appendix~\ref{sec:AppendixPSG111}). These are simply the 0- and $\pi$-flux states that were already present in the absence of a field as illustrated in Fig.~\ref{fig:synopsis}(a) and (b), respectively. 

Next, under a $[001]$ field, Zeeman couplings on each sublattice are $-\frac{h}{\sqrt{3}}(1,-1, -1, 1)$. The corresponding symmetry group is generated by translation, an improper rotation $\bar{C}_4=\bar{C}_6^2 S \bar{C}_6^{-1}$ along with the inversion symmetry. The allowed symmetric states have flux $(n_1\pi, n_1\pi, n_1\pi, n_1\pi)$ with $n_1\in\{0, 1\}$. These are once again the $0$- and $\pi$-flux states. 

Finally, under a $[110]$ field, Zeeman interactions on the four tetrahedron sites are $-\sqrt{\frac{2}{3}}h(1,0,0,-1)$. Here, the remaining spatial symmetries are generated by a mirror reflection $\sigma= S \bar{C}_6^3$, the inversion symmetry $I$, and lattice translations $T_i$. It turns out that under this magnetic field, all possible flux configurations for symmetric states are $(n_1\pi, n_2\pi, n_2\pi, n_1\pi)$, where $n_1, n_2\in \{0, 1\}$. Here, we see two novel staggered flux spin liquids are allowed where half of the hexagonal plaquettes are 0-flux and the other half, $\pi$-flux, as illustrated in Fig.~\ref{fig:synopsis}(c) and (d). Now that we have identified novel symmetric QSLs, it is then of interest to see whether there exist regions in the parameter space where these states can be stable by computing the phase diagrams.


\section{Phase Diagram \label{sec:phase_diagram}}

The phase diagram is computed by comparing the ground state energy of all symmetric Ans\"atze at different values of the transverse coupling and magnetic field strength. The resulting phase diagrams of the XXZ model~\eqref{eq:H_XXZ} for a field along the [110], [111], and [001] directions are shown in Fig.~\ref{fig:synopsis}(1a), (1b), and (1c), respectively. At sufficiently large field strength, a polarized paramagnetic phase is obtained for all field directions~\cite{li2017symmetry}. Note that we leave the nature of the polarized paramagnetic state unspecified since we are only interested here in the deconfined phase, which GMFT is better suited to describe.

It is worth noting that Yan, Sanders, Castelnovo, and Nevidomskyy~\cite{yan2023experimentally} studied a similar model~\eqref{eq:H_XXZ} perturbatively by projecting interactions into the spin ice (i.e., chargeless) manifold. There, they observe a frustrated flux phase under a finite [111] field where the flux over some hexagonal plaquettes can take on continuous (non-$\pi$-multiple) values. However, as discussed in Sec.~\ref{sec:magenabled}, PSG analysis only admits the 0-flux and $\pi$-flux phase for this field direction. As such, a phase with continuous fluxes would have to break inversion symmetry, as argued in Appendix~\ref{sec:AppendixPSG0}, and is not considered in our analysis.


\section{Experimental Signatures\label{sec:experimentalsig}}

The phase diagrams in Fig.~\ref{fig:synopsis}(2a)-(2c) show that different deconfined phases are stabilized depending on the magnitude and direction of the magnetic field. An important question is how to distinguish these phases experimentally, especially in neutron scattering experiments. In the following, we will investigate the evolution of the neutron scattering signatures upon the application of a magnetic field starting from $J_\pm=0.03$, $J_\pm=-0.03$, and $J_\pm=-0.3$ (in the unit of $J_{yy}=1$). The choice of $J_\pm=-0.3$ is based on recent estimates for Ce$_2$Zr$_2$O$_7$ and can offer a meaningful comparison to future experiments. Along $J_\pm=-0.03$, we can investigate spinon dynamics during phase transitions from $\pi$ into $0$-flux or $(0,\pi,\pi,0)$ states. Finally, $J_\pm=0.03$ probes how 0-flux DO-QSI evolves under magnetic fields.

\begin{figure}
    \centering
    \includegraphics[width=\linewidth]{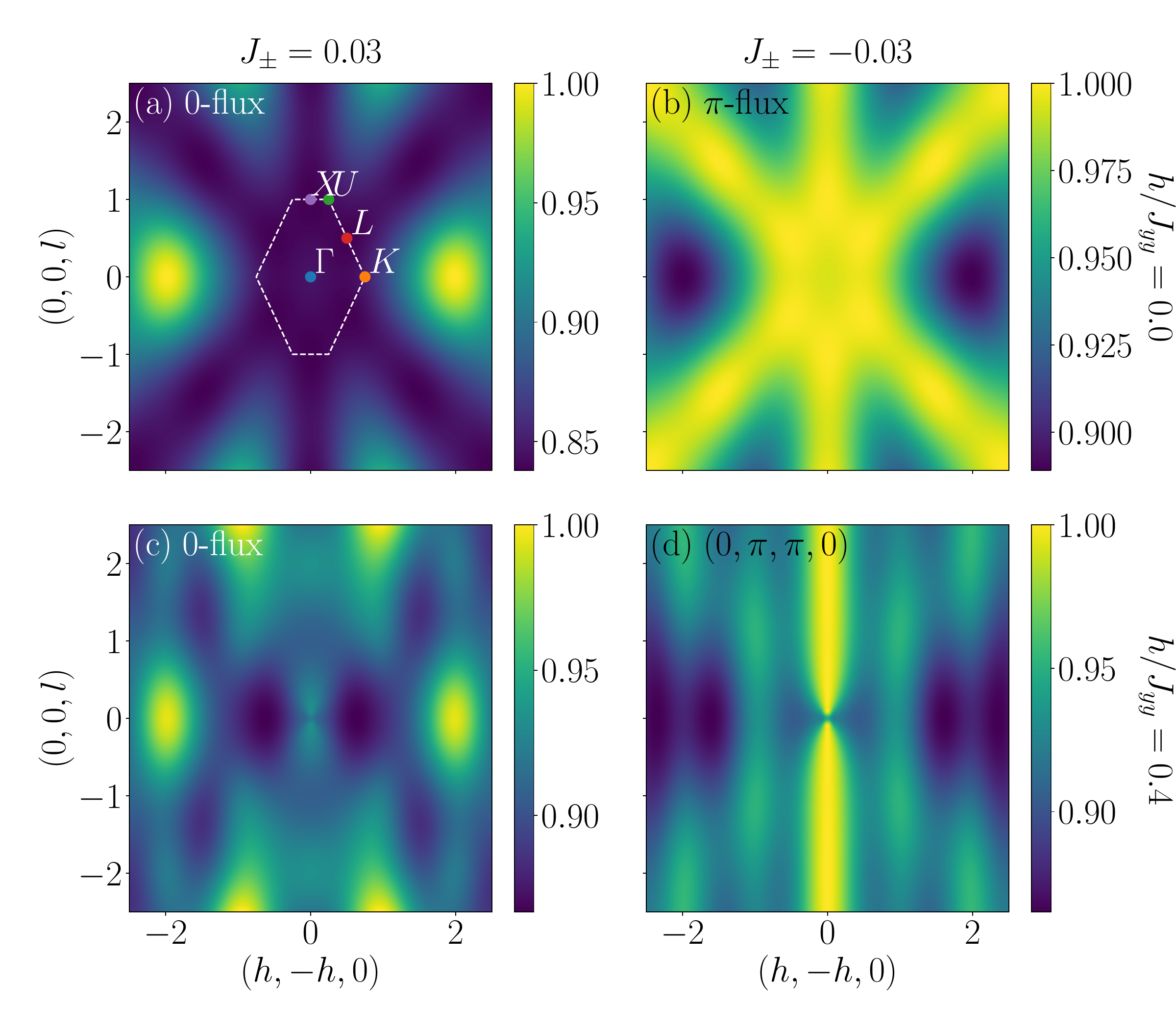}
    \caption{Static spin structure factors in the global frame $\mathcal{S}^{zz}(\mathbf{q})$ with a field in the $[110]$ direction along the $(h,-h,l)$ plane for a transverse coupling of (a), (c) $J_\pm/J_{yy} = 0.03$ and  (b), (d) $J_\pm/J_{yy} = -0.03$ at a field strength of (a)-(b) $h/J_{yy}=0$ and (c)-(d) $h/J_{yy}=0.4$. (d) is in the $(0,\pi,\pi,0)$ phase. The First Brillouin zone is highlighted in white, and high symmetry points are labeled accordingly.}.
    \label{fig:SSSF110}
\end{figure}

Since both $\tau^{x}$ and $\tau^{y}$ have an underlying octupolar magnetic charge distribution (see Eq.~\eqref{eq:gamma5gamma6}), they have a vanishing magnetic form factor for small momentum transfer~\cite{yahne2022dipolar, sibille2020quantum}. As a result, neutron scattering (for small momentum transfer) only probes correlations between $\tau^z$ ($\tau^z=S^{z}$ here since $\theta=0$). The neutron scattering cross-section is then proportional to \begin{align}
    \frac{d^2\sigma}{d\Omega d\omega} \propto \sum_{\mu,\nu}\mathcal{S}_{\mu\nu}^{zz}(\mathbf{q},\omega) = \mathcal{S}^{zz}(\mathbf{q},\omega), \label{eq:neutron}
\end{align}
where  
\begin{equation}
    \mathcal{S}_{\mu\nu}^{zz}(\mathbf{q},\omega) =\left(\hat{\mathbf{z}}_\mu \cdot \hat{\mathbf{z}}_\nu - \frac{(\hat{\mathbf{z}}_\mu\cdot \mathbf{q})(\hat{\mathbf{z}}_\nu\cdot \mathbf{q})}{|\mathbf{q}|^2} \right)  \mathcal{S}_{\mathrm{LF},\mu\nu}^{zz} (\mathbf{q},\omega) \label{eq:szzmunu},
\end{equation}
and $\mathcal{S}_{\mathrm{LF},\mu\nu}^{\alpha\beta} (\mathbf{q},\omega)$ is the dynamical spin-spin correlation in the local sublattice-dependent frame 
\begin{equation}\label{eq:SSF}
     \mathcal{S}^{\alpha\beta}_{\mathrm{LF}, \mu\nu}(\mathbf{q},\omega)  = \sum_{\mathbf{R}_\mu, \mathbf{R}_\nu} e^{i\mathbf{q}(\mathbf{R}_\mu-\mathbf{R}_\nu)} \int dt e^{i\omega t}  \langle S^\alpha_{\mathbf{R}_\mu}(t)S^\beta_{\mathbf{R}_\nu}(0)\rangle.
\end{equation}
The above considerations and the mapping to emergent quantum electrodynamics defined in Eq.~\eqref{eq:mapping_pseudospin_eQED} implies that neutron scattering in the regime of interest ($J_{yy}>J_{xx},J_{zz}$) only probes the spinons $\mathcal{S}^{zz}\sim\expval{\Phi^\dagger \Phi \Phi^\dagger \Phi}$, but not the photon $\expval{S^y S^y}\sim \expval{E E}$. Note that, considering the above, whenever spin correlations are discussed below, it is implicitly assumed that these are correlations between $S^{z}$ spin components (i.e., correlations in the $\ket{S^{z}=\pm}$ basis). 

\subsection{Static Spin Structure Factor \label{sec:SSSF}}

Let us first examine the equal-time or static spin structure factor (SSSF), which corresponds to energy-integrated results in neutron scattering experiments $\mathcal{S}^{zz}(\mathbf{q}):=\int d\omega \mathcal{S}^{zz}(\mathbf{q},\omega)$. The experimentally relevant results for $J_\pm=-0.3$ are presented in Fig.~\ref{fig:synopsis}(2) and the results for $J_\pm=0.03$ and $J_\pm=-0.03$ are presented in Fig.~\ref{fig:SSSF110}, \ref{fig:SSSF111}, and \ref{fig:SSSF001} for field directions [110], [111], and [001], respectively. These results showcase unique signatures, most notably under a [110] field. Moreover, these results also shed light on the nature of the staggered flux phase. Let us examine them in close detail below. 

\subsubsection{\texorpdfstring{$\mathbf{B}\parallel[110]$}{\texttwoinferior}}\label{sec:h110}

Under a $[110]$ field, two of the four sublattices (1 and 2) of the tetrahedron are completely decoupled, whereas the other two (0 and 3) are strongly coupled to the field. One can, therefore, break the system down into two types of $1d$ chains: the coupled chain $\alpha$ and the decoupled chain $\beta$ as shown in Fig.~\ref{fig:pyrochlore}(d). At large fields, the interchain coupling between the $\alpha$ and $\beta$ chains is approximately zero (compared to the dominant Zeeman term) as the pseudospins on the $\alpha$ chain are strongly polarized, and the system is effectively quasi-1-dimensional with $\alpha$ and $\beta$ chains stacked along the $[0,0,l]$ direction.

Fig.~\ref{fig:SSSF110} shows the evolution of SSSF under a [110] magnetic field. One immediate striking feature is that, regardless of $J_\pm$, there is a distinctive emergent ``pinch point" at the $\Gamma$ point with an overall enhanced rod-like intensity along $(0,0,l)$ as we increase the magnetic field. By examining the appropriate sublattice contribution to the SSSF signature $\mathcal{S}^{zz}_{\mu\nu}$, we attribute this effect to the increasingly polarized $\alpha$ chains, as reasoned in Appendix~\ref{sec:sublatticeSSSF110}. On the other hand, the $\beta$ chains contribute to intensities forming the typical snowflake patterns in the $(h,-h,l)$ scattering planes with opposite intensities for 0-flux and $\pi$-flux QSI that have already been identified in previous studies~\cite{desrochers2023spectroscopic, Hosoi2022Uncovering} in the absence of a magnetic field. Since the $\beta$ chains are decoupled from the magnetic fields, these snowflake patterns with opposite intensities remain as we increase the magnetic fields, although their shapes are distorted by the emergence of stripe-like patterns.

Another distinctive feature under this magnetic field is the development of stripes patterns along the $(0,0,l)$ direction as shown in Fig.~\ref{fig:SSSF110}(c)-(d) and in Fig.~\ref{fig:synopsis2}(2d). Different from the rod-like motifs (i.e., snowflake pattern) seen in Fig.~\ref{fig:SSSF110}(b), which have been attributed to be artifacts of the transverse projector~\cite{castelnovo2019rod}, here the stripes are due to loss of correlation along $(0,0,l)$. Indeed, the correlation between the chains decreases as the $\alpha$ chains get progressively more polarized. The positions of the stripes reveal the underlying physics of the different phases. For $\pi$-flux QSI at $J_\pm=-0.3$, the stripe pattern develops intensities at $X$ and $L$ points. This is due to a strong antiferromagnetic inter-$\alpha$-chain correlation, as reasoned in Appendix~\ref{sec:sublatticeSSSF110}. In contrast, 0-flux QSI and $(0,\pi,\pi,0)$ have instead ferromagnetic inter-$\alpha$-chain correlations, as seen by stripe features positioned at the $\Gamma$ points this time. 
\begin{figure}
    \centering
    \includegraphics[width=\linewidth]{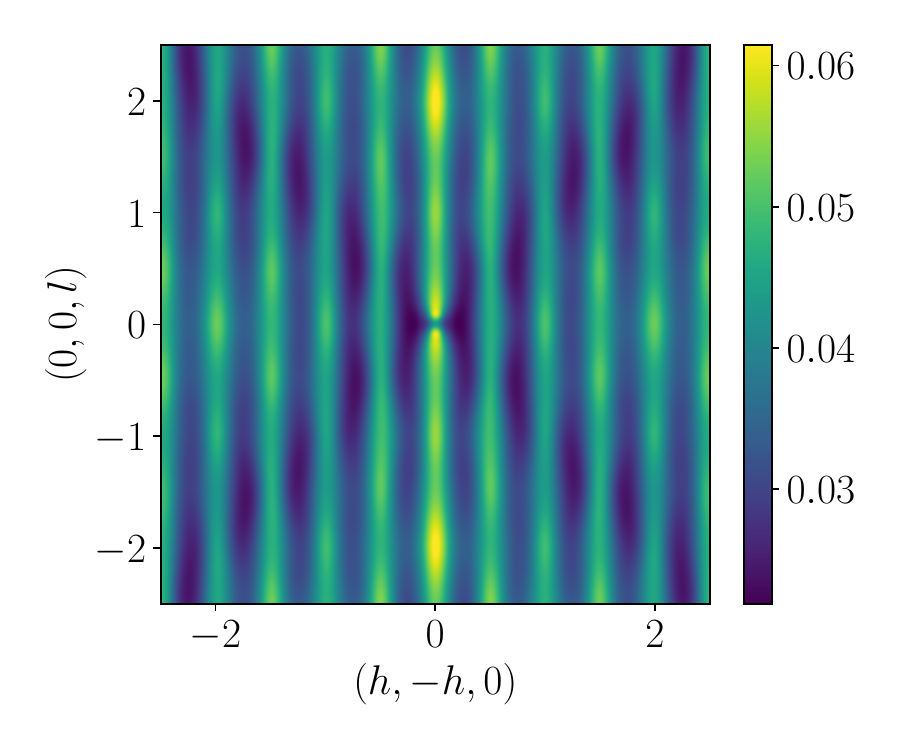}
    \caption{Static spin structure factor for $J_\pm=-0.3J_{yy}$, under a [110] field of $h=0.1J_{yy}$, after subtracting the corresponding zero field results.}
    \label{fig:subtracted}
\end{figure}

We would like to compare our results with a recent experimental study on Ce$_2$Zr$_2$O$_7$~\cite{smith2023quantum}, where magnetic field dependence of the equal-time structure factor is studied in neutron scattering. Given that an experimentally determined set of exchange parameters is $(J_{xx},J_{yy},J_{zz}) = (0.062 {\rm meV}, 0.063 {\rm meV}, 0.011 {\rm meV})$ and $J_{\pm} = - (J_{xx} + J_{zz})/4$,
we get $J_{\pm}/J_{yy} \approx - 0.3$. Hence, our phase diagram for $J_{\pm}/J_{yy} = - 0.3$ is the most relevant to the experiment. 
Note that, for this parameter, the $\pi$-flux QSI is not stable beyond $h_c \approx 0.25 J_{yy}$ in the [110] field as shown in Fig.~\ref{fig:synopsis2}(1a). 
The smallest magnetic field used in Ref.~\cite{smith2023quantum} was $h=0.35\mathrm{T}\approx  0.72J_{yy}$.
This means the system is already in the polarized state (spinons are already condensed) for the smallest field used in the neutron scattering experiment. In the experiment, the magnetic field dependence of the equal-time structure factor is demonstrated by subtracting the zero field results from that of a finite [110] field. 
For meaningful comparisons, we perform a similar analysis by subtracting the zero field results from the $h=0.1J_{yy}$ case as shown in Fig.~\ref{fig:subtracted}. Here, the system is in the $\pi$-flux QSI phase at $h=0.1J_{yy}$. 
Note that the stripe patterns, as well as the pronounced $(0,0,l)$ rod with a ``pinch point" at $\Gamma$ are visible in both experimental data at $h=0.35\mathrm{T}\approx0.72J_{yy}$ and the theoretical results at $h=0.1J_{yy}$. This is mainly because these features are signatures of the $\alpha$ chains, which should remain visible regardless of the presence/absence of spinon condensation. Indeed, the theoretical results at $h=0.1J_{yy}$ and experimental results at $h=0.35\mathrm{T}\approx0.72J_{yy}$ show many similarities. Namely, the rod-like features at $(0, 0, l)$, $(\pm1/2, \mp1/2, l)$, and $(\pm1, \mp1, l)$ are captured along with peak intensities at $(\pm2,\mp2,0)$ and $(0,0,\pm2)$. It will be interesting to perform the neutron scattering at smaller magnetic fields for better comparison with theoretical results.

\begin{figure}
    \centering
    \includegraphics[width=\linewidth]{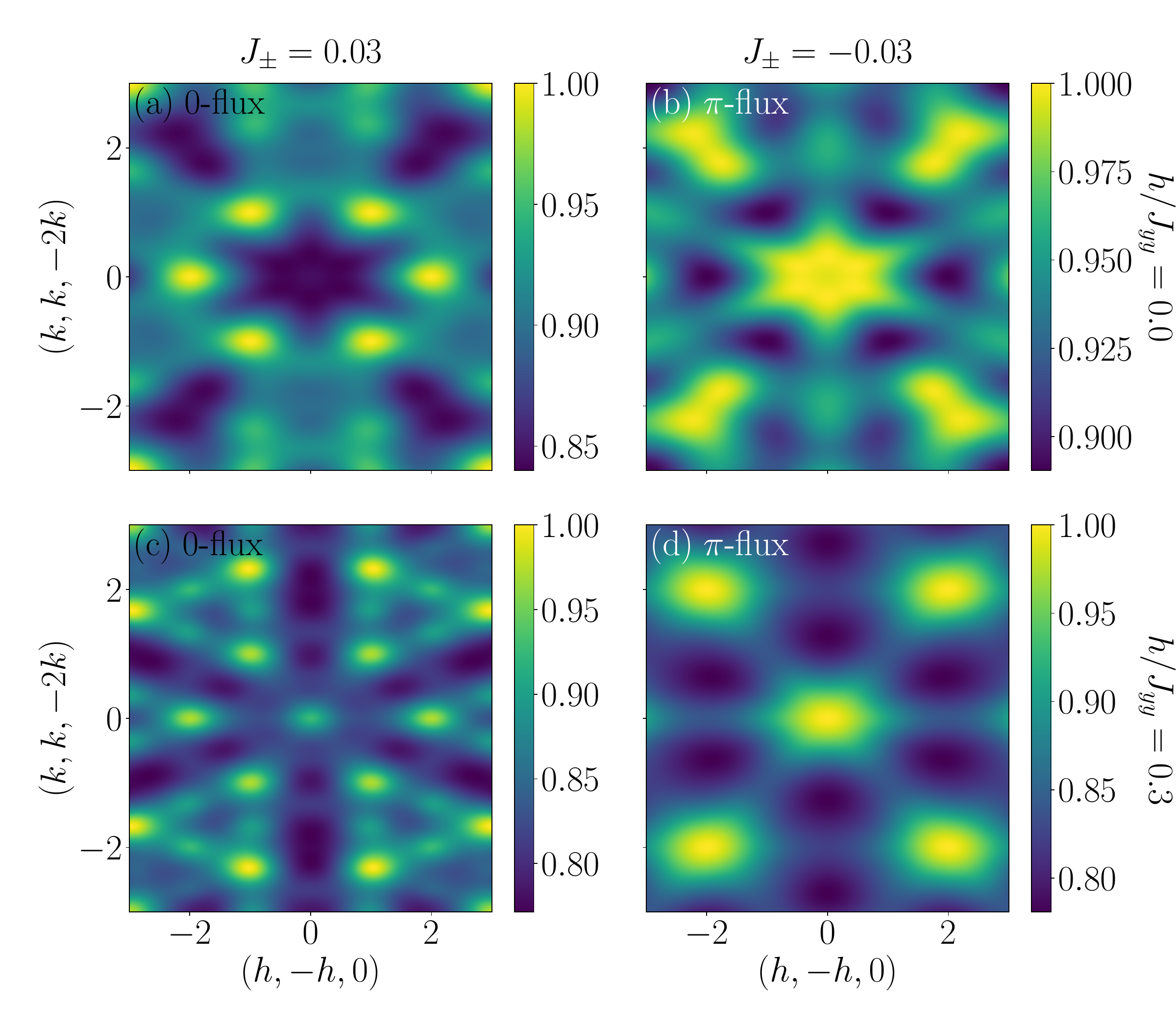}
    \caption{
    Static spin structure factors in the global frame $\mathcal{S}^{zz}(\mathbf{q})$ with a field in the $[111]$ direction along the $(h+k,-h+k,-2k)$ plane for a transverse coupling of (a), (c) $J_\pm/J_{yy} = 0.03$ and  (b), (d) $J_\pm/J_{yy} = -0.03$ at a field strength of (a)-(b) $h/J_{yy}=0$ and (c)-(d) $h/J_{yy}=0.3$.}
    \label{fig:SSSF111}
\end{figure}

\subsubsection{The \texorpdfstring{$(0,\pi,\pi,0)$}{\texttwoinferior} Phase \label{sec:0pp0}}
The nature of $(0,\pi,\pi,0)$ can be elucidated from the results above. 
Namely, this is a state where the intra-$\beta$-chain correlation remains antiferromagnetic as shown in Fig.~\ref{fig:sublattice110_3}(d), whereas the inter-$\alpha$-chain correlations are ferromagnetic as shown in Fig.~\ref{fig:sublattice110_2}(c). This contrasts the 0-flux ($\pi$-flux) phase where intra-$\beta$-chain correlations and inter-$\alpha$-chains are both ferromagnetic (antiferromagnetic). The $[110]$ magnetic field energetically favors ferromagnetic correlations between the $\alpha$ chains by polarizing the spins, while a negative transverse coupling $J_{\pm}$ favors antiferromagnetic intra-$\beta$-chain correlations. As such, at large enough strength, it becomes more energetically favorable for the system to adopt the $(0,\pi,\pi,0)$ state with its mixed correlations instead of 0- or $\pi$-flux QSI.

\subsubsection{\texorpdfstring{$\mathbf{B}\parallel[111]$}{\texttwoinferior}\label{sec:h111}} 

Under a [111] field, the coupling strengths on sublattices 1, 2, and 3 are the same, whereas, at sublattice 0, the Zeeman coupling is stronger by a factor of 3. As such, we can effectively group the system into layers of sparse triangular and Kagome planes, as shown in Fig.~\ref{fig:pyrochlore}(f). 

\begin{figure}
    \centering
     \includegraphics[width=\linewidth]{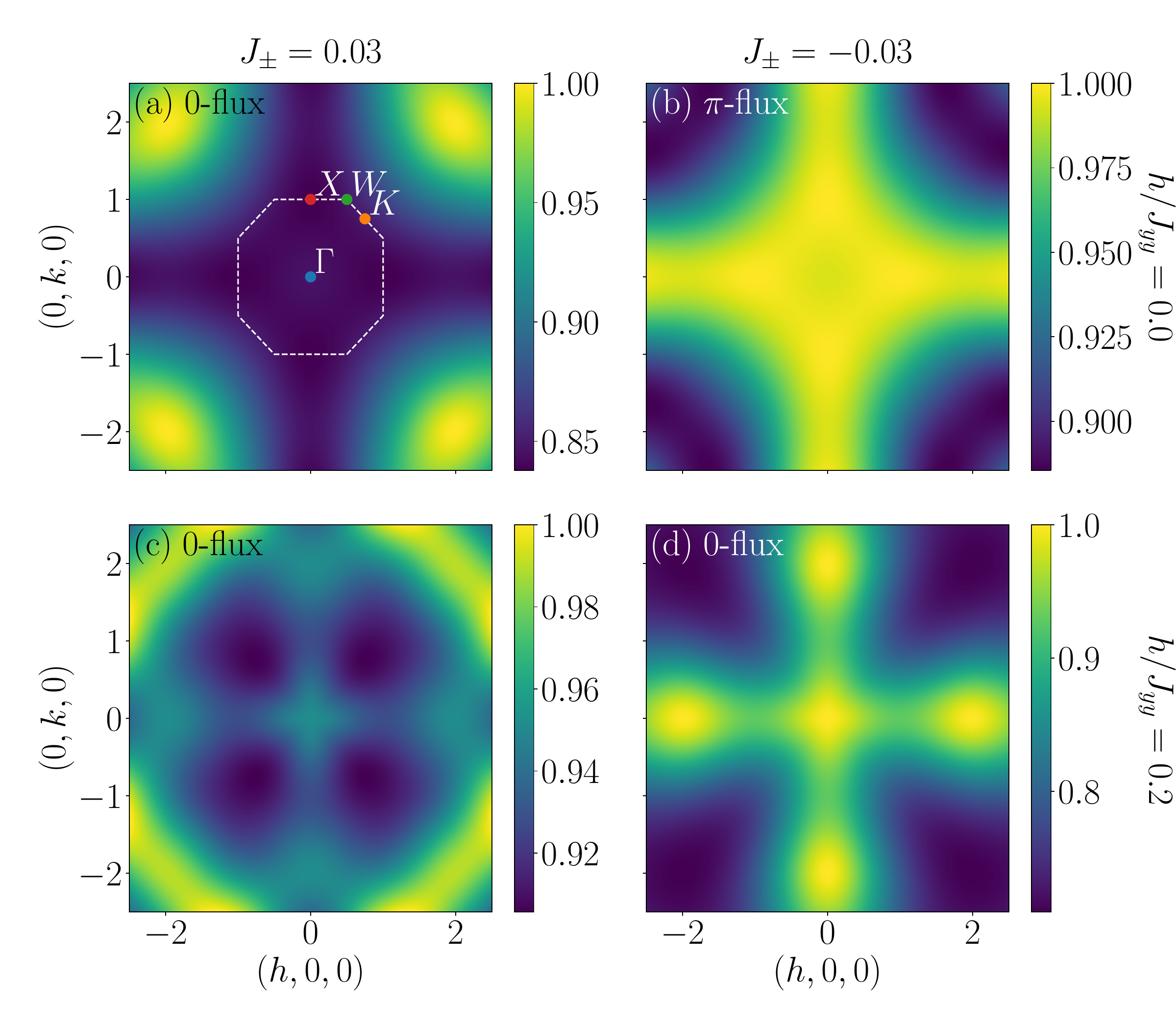}
    \caption{
    Static spin structure factors in the global frame $\mathcal{S}^{zz}(\mathbf{q})$ with a field in the $[001]$ direction along the $(h,k,0)$ plane for a transverse coupling of (a), (c) $J_\pm/J_{yy} = 0.03$ and  (b), (d) $J_\pm/J_{yy} = -0.03$ at a field strength of (a)-(b) $h/J_{yy}=0$ and (c)-(d) $h/J_{yy}=0.2$. (d) is in the 0-flux phase despite $J_\pm<0$. The first Brillouin zone is highlighted in white, and high symmetry points are labeled accordingly.}
    \label{fig:SSSF001}
\end{figure}
\begin{figure*}
    \centering
    \includegraphics[width=\linewidth]{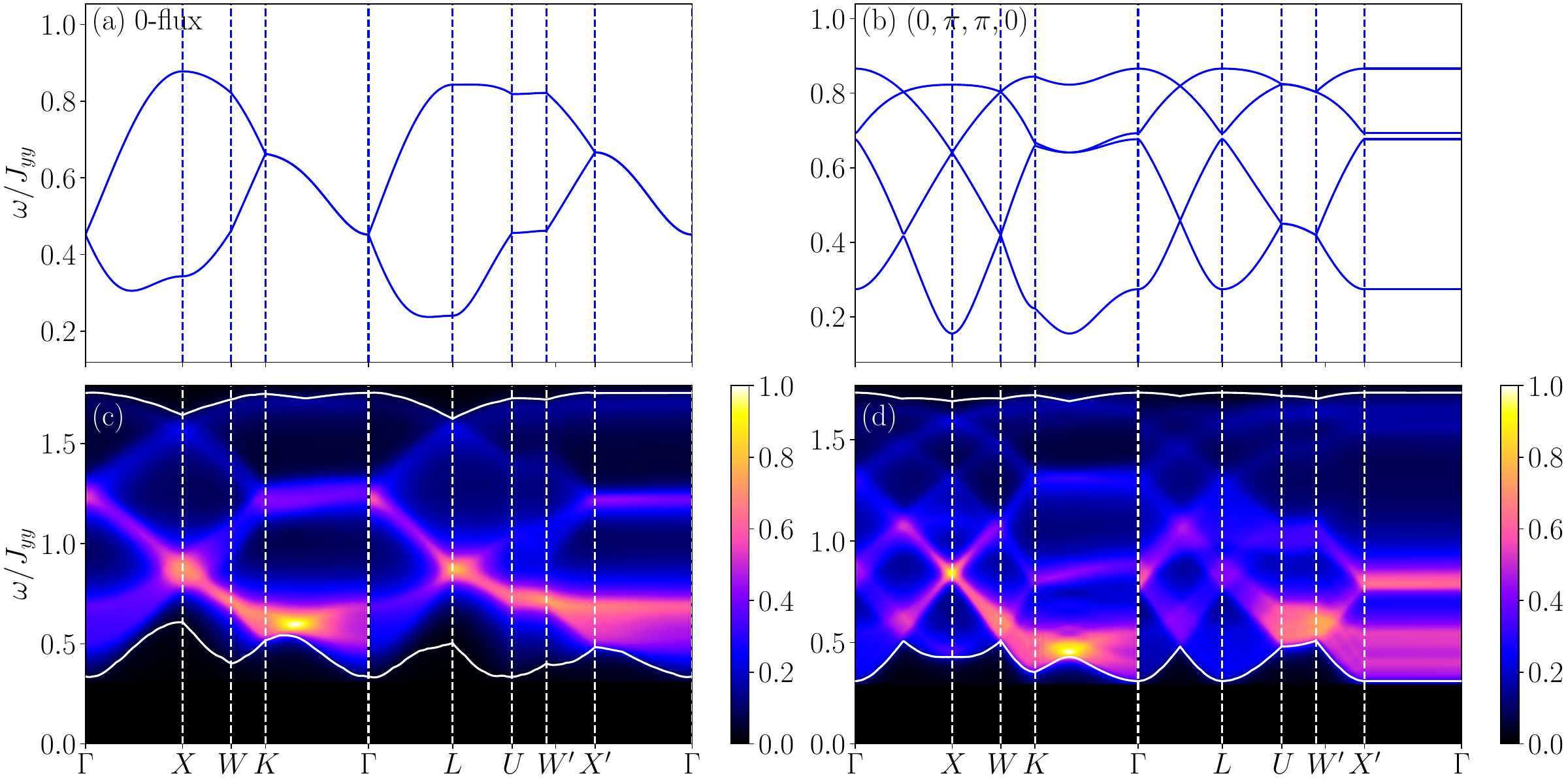}
    \caption{
    (a)-(b) Spinon dispersion and (c)-(d) dynamical spin structure factor in the global frame $\mathcal{S}^{zz}(\mathbf{q},\omega)$ under a $[110]$ for (a), (c) $J_\pm/J_{yy}=0.03$, $h/J_{yy}=0.4$ and (b), (d)$J_\pm/J_{yy}=-0.03$, $h/J_{yy}=0.4$. The upper and lower edges of the two-spinon continuum are denoted by white lines.}
    \label{fig:DSSF110}
\end{figure*}

The $\mathcal{S}^{zz}(\mathbf{q})$ in Fig.~\ref{fig:synopsis}(2e) for the $\pi$-flux QSI at $J_\pm=-0.3$ resembles that reported for Kagome spin ice systems. A quantum Monte Carlo study was carried out on an effective spin-1/2 system with dominant coupling $J_{zz}$ under a [111] field that couples to the local $S^z$ pseudospins only~\cite{bojesen2017Quantum}. The prediction for SSSF was computed, as shown in Fig.~4(b) in Ref.~\cite{bojesen2017Quantum}. We can see that this closely resembles the results of this work. In particular, we manage to capture the same peak intensities at $(\frac{2}{3}, -\frac{2}{3}, 0)$ (and symmetry-related points) as well as intensities along $(h,0,-h)$. 

This is a remarkable observation, considering that the microscopic physics of such a system is completely different from the DO case we are studying. For the effective spin-1/2 QSI, the 2-in-2-out configuration in the $\ket{S^z=\pm}$ basis~\cite{onoda2011quantum} is favored at zero field (i.e., the spin ice rules are respected). For any finite field along [111], it then becomes energetically favorable for the system to select a subset of states in the degenerate spin ice manifold where the $S^z$ components of pseudospins on the triangular plane align with the [111] magnetic field. The spins of the Kagome plane then follow the 2-in-1-out Kagome ice rule to preserve the spin ice rules for every tetrahedron. This results in the signatures described above.

In contrast, the $\pi$-flux DO-QSI state of interest approximately respects the spin ice rules in the $\ket{S^y=\pm}$ basis. It is then intriguing how, despite their markedly distinct underlying physics, the $\pi$-flux DO-QSI and effective spin-1/2 QSI could feature similar SSSF under a [111] field.
This can be reasoned by noting that much of the same kagome ice physics described above is present in a much more subtle way for $\pi$-flux DO-QSI. As reasoned in Ref.~\cite{desrochers2023spectroscopic}, the $\pi$-flux QSI state at zero fields is a superposition of all tetrahedron configurations in the $\ket{S^z=\pm}$ basis where 2-in-2-out configurations are favored over all-in-all-out. As such, when $S^z$ on the triangular plane is pinned by a finite magnetic field, $\pi$-flux QSI will favor 2-in-1-out in the Kagome planes and display Kagome-ice-like correlations (see Appendix \ref{sec:sublatticeSSSF111} for detailed argument). 

On the other hand, when $J_\pm > 0$ (i.e., 0-flux), AIAO configurations are favored instead of the 2-in-1-out Kagome ice rule in the $\ket{S^z=\pm}$ basis. This manifests in the opposite SSSF patterns at weak fields as shown in Fig.~\ref{fig:SSSF111}(c)-(d) between the 0-flux and $\pi$-flux phase.

\subsubsection{\texorpdfstring{$\mathbf{B}\parallel[001]$}{\texttwoinferior}\label{sec:h001}}

At large fields, the polarized paramagnetic state under the [001] field is 2-in-2-out in the $|S^z=\pm\rangle$ basis with configurations shown in Fig.~\ref{fig:pyrochlore}(d). For $\pi$-flux DO-QSI, as we increase the magnetic field, the system simply selects a specific superposition of states that follows such a 2-in-2-out configuration. Therefore, the snowflake pattern seen in $\pi$-flux DO-QSI at zero field~\cite{ sibille2018experimental,desrochers2023spectroscopic, Hosoi2022Uncovering} survives as we increase the magnetic field. In the $(h,k,0)$ scattering plane, the snowflake pattern is seen in Fig.~\ref{fig:SSSF001}(d) and Fig.~\ref{fig:synopsis}(2f) as a cross along the $(h,0,0)$ and $(0,k,0)$ direction. On the other hand, for 0-flux DO-QSI, at large enough fields, the system eventually favors the 2-in-2-out configuration (instead of all-in-all-out at zero fields). As a result, correlations along the $(h,0,0)$ and $(0,k,0)$ cross go from weak to strong as the field is increased (see in Fig.~\ref{fig:SSSF001}(c)).

\subsection{Dynamical Spin Structure Factor\label{sec:DSSF}}
\begin{figure*}
    \centering
    \includegraphics[width=\linewidth]{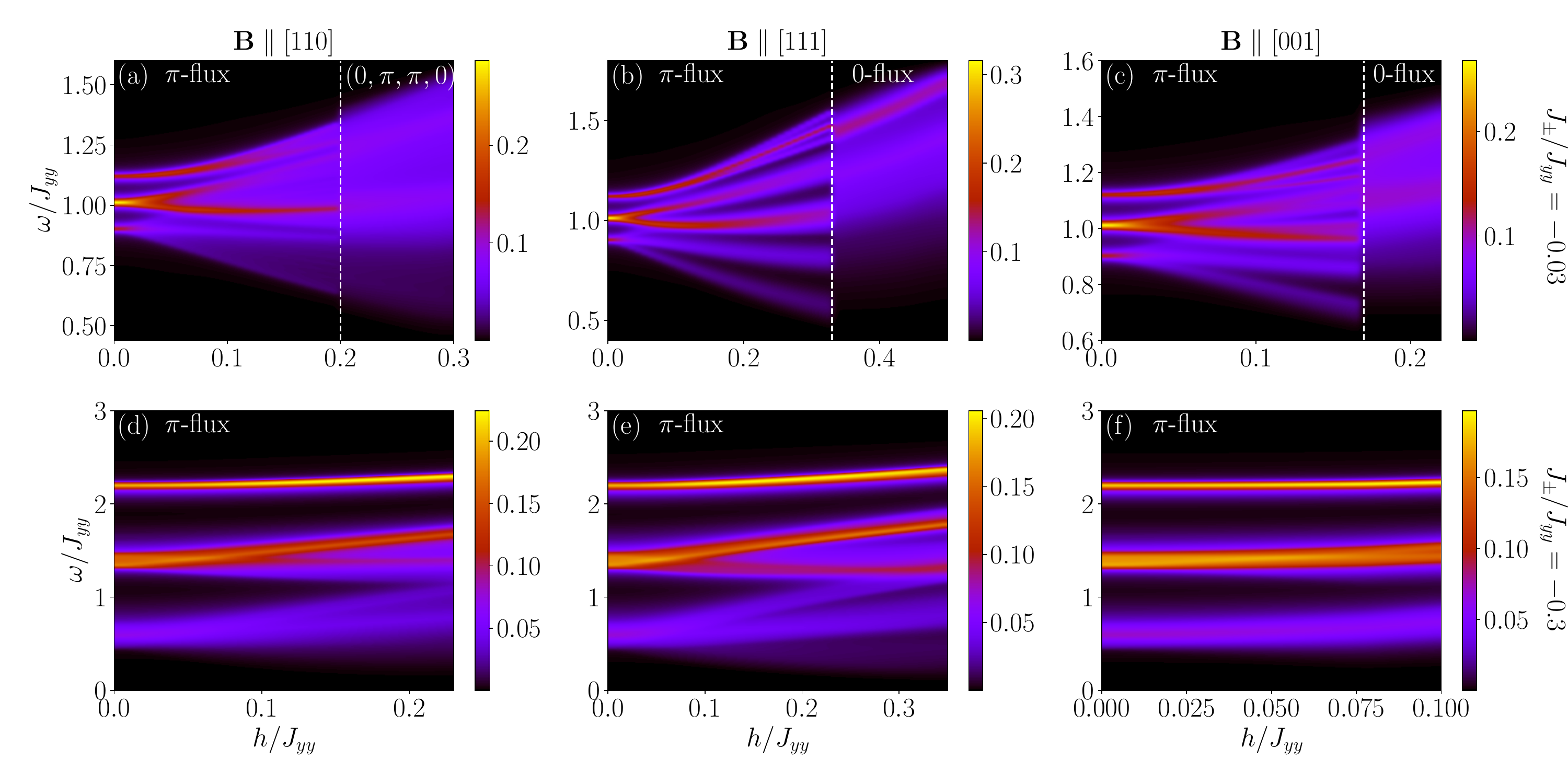}
     \caption{Two spinon density of states under (a), (d) the [110] field, (b), (e) the [111] field, and (c), (f) the [001] field at varying magnetic field strength for (a)-(c) $J_\pm/J_{yy}=-0.03$ and (d)-(f) $J_\pm/J_{yy}=-0.3$. The first-order transitions are denoted by the white dashed lines.}
    \label{fig:2SpinonDOS111Jpm-0.03}
\end{figure*}

\begin{figure}
    \centering
    \includegraphics[width=\linewidth]{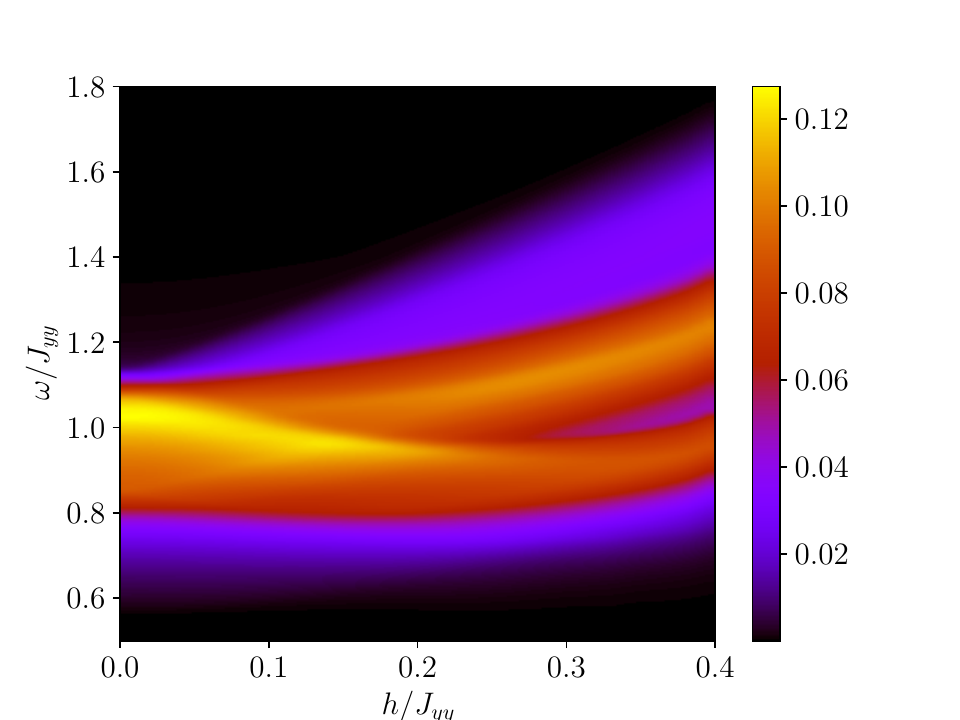}
    \caption{Two spinon density of states for 0-flux QSI with $J_\pm/J_{yy}=0.03$ under a [111] field as a function of magnetic field strength.}
    \label{fig:2SDOS_Jpm=0.03}
\end{figure}

We now provide predictions for the DSSF~\eqref{eq:szzmunu} and spinon dispersions. Due to finite magnetic fields, some pyrochlore symmetry points in the Brillouin are inequivalent. As such, let us define a path as shown in Fig.~\ref{fig:DSSF110} given in reciprocal lattice unit under cubic coordinates, with $\Gamma=(0,0,0)$, $X=(1,0,0)$, $W=(1,-1/2,0)$, $K=(3/4,-3/4,0)$, $L=(1/2,1/2,1/2)$, $U=(1/4, 1/4, 1)$, $W'=(0,1/2,1)$, $X'=(0,0,1)$. This path probes along all scattering planes discussed in Sec. \ref{sec:SSSF}. 

Without a field, the A and B diamond sublattices are completely decoupled for the XXZ model (see Eq.~\eqref{eq:H_parton}), resulting in degenerate spinon bands for the two sublattices.
More specifically, 0-flux has a single non-degenerate spinon band~\cite{savary2012coulombic}, whereas $\pi$-flux has two~\cite{lee2012generic}.
Introducing the Zeeman term allows spinon hopping between the two diamond sublattices. As a result, the degeneracy between the A and B sublattice is lifted. The number of non-degenerate bands is accordingly doubled compared to the zero field case, as shown in Fig.~\ref{fig:DSSF110}(a)-(b). This degeneracy lifting results in drastically different DSSF for 0- and $\pi$-flux QSI compared to the zero field case. 

For 0-flux QSI, under finite fields $\mathbf{B}\parallel[111]$, the momentum-integrated DSSF develops two well-resolved local maxima. This is most readily captured by looking at the evolution of two-spinon density of states (DOS) in Fig.~\ref{fig:2SDOS_Jpm=0.03}, which correspond to the location of the peaks in DSSF upon momentum integration $\rho^{(2)}(\omega)=\sum_{\mathbf{k},\mathbf{k}',\sigma,\lambda}\delta(\omega-\epsilon_\sigma(\mathbf{k})-\epsilon_\lambda(\mathbf{k}'))/N^2\sim \int d\mathbf{q} \mathcal{S}^{zz}(\mathbf{q},\omega)$. Indeed, we see that the two spinon DOS has two well-resolved peaks at large magnetic fields.

On the other hand, for the $\pi$-flux phase where $J_\pm=-0.03$, the spinon dispersion is composed of two flat bands at zero field. Under a finite magnetic field $\mathbf{B}\parallel[110]$, the $\pi$-flux state dispersion remains mostly flat. As such, the two flat spinon bands split into four. This is also true for all other magnetic field directions, as shown in Fig.~\ref{fig:DSSF111}(b) and Fig.~\ref{fig:DSSF001}(b). Because of the flatness in the spinon bands, there still exist well-resolved peaks in the DSSF upon momentum integration. Due to the splitting of these bands, the 3-peak signature at zero field splits into multiple peaks for all field directions. Furthermore, the ways they evolve under different field directions are drastically different. Again, we can compute the evolution of the two spinon DOS with a magnetic field in Fig.~\ref{fig:2SpinonDOS111Jpm-0.03}(a)-(c). Under a [110] field, the 3-peak structure of $\pi$-flux QSI first splits into five peaks at low fields and then further splits into roughly seven peaks just before phase transition. Similarly, under a [001] field, the 3-peak structure first splits into 5 peaks and then further splits into nine peaks. On the other hand, under a [111] field, the 3-peak structure splits into five peaks, which remains well-resolved just before condensation with some additional splittings of the topmost band as shown in Fig.~\ref{fig:2SDOS_Jpm=0.03}(b).

For a DO Ce-based compound where $J_\pm\approx -0.3$, since $h_c$ is weak, the effect of splitting is small even at $h_c$ for each respective magnetic field direction. Under a [110] field, the three peaks split into five where the splitting is on the scale of $\approx 0.25J_{yy}$ at $h=0.23J_{yy}$ as shown in Fig.~\ref{fig:2SpinonDOS111Jpm-0.03}(d). We also see the splitting of three peaks into five peaks under a [111] field with a much more drastic splitting on the scale of $\approx0.5J_{yy}$ at $h_c\approx0.3 J_{yy}$ in Fig.~\ref{fig:2SpinonDOS111Jpm-0.03}(e). Finally, there is virtually no splitting under the [001] field in Fig.~\ref{fig:2SpinonDOS111Jpm-0.03}(f). The evolution of DSSF under these three magnetic field directions makes up a unique experimental profile of the $\pi$-flux DO-QSI due to a strong dependence on the field orientations. However, these fine features may be difficult to resolve experimentally, given that the dominant microscopic exchange parameters have shown to be on the scale of $0.06$meV.

It is worth noting that the enhanced spectral periodicity seen in $\pi$-flux QSI~\cite{essin2014spectroscopic, chen2017spectral, desrochers2023symmetry} is still present under finite fields. Because of the symmetry fractionalization of the spinons~\cite{chen2017spectral}, the $\pi$-flux phase has enhanced periodicity such that the two-spinon DOS is equal at the $\mathbf{q}$, $\mathbf{q}+2\pi(1,0,0)$, $\mathbf{q}+2\pi(0,1,0)$, and $\mathbf{q}+2\pi(0,0,1)$ points (in cubic coordinates). As shown in Fig.~\ref{fig:synopsis2}(3a)-(3c), the upper and lower edges of the twos-spinon continuum are the same at the $\Gamma$, $X$, $X'$, and $L$ points~\cite{yao2020pyrochlore}. On the other hand, due to the staggered flux configuration of $(0,\pi,\pi,0)$, the enhanced periodicity is modified for the state such that the two-spinon DOS is now equal at $\mathbf{q}$, $\mathbf{q}+\pi(1,1-1)$ and $\mathbf{q}+2\pi(0,0,1)$, as shown in Fig.~\ref{fig:DSSF110}(d) where the two-spinon continuum edges are the same at $\Gamma$, $X'$, and $L$. See detailed derivations in Appendix~\ref{sec:enhancedperiod}.  


\section{Discussion \label{sec:discussion}}

In this work, we used the PSG analysis for DO-QSI to identify two novel QSI phases only allowed under a magnetic field along the [110] direction beyond the 0-flux and $\pi$-flux phases present at zero field. We computed the phase diagrams using the GMFT formalism for [110], [001], [111] magnetic fields and found that one of the two novel phases is indeed stable near the Ising point up to a field strength $~h/J_{yy}\approx 1$ for the [110] magnetic field. Furthermore, we found first-order field-induced phase transitions near the Ising point from $\pi$-flux QSI into $0$-flux QSI under a [111] or [001] field or into the $(0,\pi,\pi,0)$-flux phase under a [110] field.

In addition, we highlighted unique signatures in the SSSF for various QSI phases in magnetic fields along cubic directions, providing striking experimental features for the identifications of DO-QSI. Namely, under a [110] field, SSSF shows the emergence of a stripe pattern with a pronounced rod at $(0,0,l)$ and an emergent ``pinch point" at $\Gamma$. Under a [111] field, we observe Kagome-ice-like correlations. Finally, under a [001] field, the rod-like motifs of $\pi$-flux QSI persist. Furthermore, we provided predictions for the magnetic-field dependence of DSSF, corresponding to inelastic neutron scattering experiments. There, we see the 3-peak signature found in $\pi$-flux QSI under zero field split into multiple peaks depending on the magnetic field directions. Furthermore, the magnitudes of the splitting at $h_c$ are drastically different. Such a strong dependence on the direction of magnetic fields can make up a unique experimental profile for the identifications of $\pi$-flux QSI.

In our work, the relative energy of competing QSI phases is computed using GMFT and is determined by the spinon contributions in different QSI phases. 
We note that our phase diagrams are different from the results in Ref.~\cite{yan2023experimentally}, where spinons are not included in the computation of energy for different QSI phases, and instead, the energy of the emergent photons is taken into account perturbatively near the Ising point. Since our result does not directly include the photon contributions, it appears that there is a competition between spinon and photon contributions, prompting future theoretical or numerical investigations. More discussion about this point can be found in the Appendix~\ref{sec:competition}.

It can be shown that the PSG solution for a dipolar QSI, where $J_{xx}>J_{yy},J_{zz}$, is generally different from that of the octupolar QSI, where $J_{yy}>J_{xx},J_{zz}$, under a [110] and a [001] field (see Appendix~\ref{sec:dipolar_ice}). This difference is insignificant under the XXZ model, where 
$J_{xx}>J_{yy}=J_{zz}$ or
$J_{yy}>J_{xx}=J_{zz}$,
as it results in the same experimental signatures. However, it can be important when considering the XYZ model, where $J_{xx}>J_{yy} \not=J_{zz}$ or
$J_{yy}>J_{xx} \not=J_{zz}$,
because more mean-field parameters arise in GMFT. It would be interesting to see if this can potentially provide convincing ways to distinguish between dipolar and octupolar QSIs. 


\section*{Acknowledgement}
We thank Han Yan for the helpful discussions.
This work was supported by the Natural Science and Engineering Council of Canada (NSERC) Discovery Grant No. RGPIN-2023-03296 and the Center for Quantum Materials at the University of Toronto. Computations were performed on
the Niagara cluster, which SciNet hosts in partnership
with the Digital Research Alliance of Canada. F.D. is supported by the Vanier Canada Graduate Scholarship. Z.Z. is supported by the Canada Graduate Scholarship (CGSM).


\appendix

\section{Pyrochlore Lattice \label{sec:AppendixPyrochlore}}
Sublattice indexed pyrochlore lattice (SIPC), $\mathbf{R}_\mu$, and sublattice indexed diamond lattice (SIDC), $\mathbf{r}_\alpha$ are used in the main text to label the pyrochlore sites and the parent diamond lattice sites, respectively. They are related through:
\begin{align}
     \mathbf{R}_{\mu}  &= r_1 \mathbf{e}_1 + r_2 \mathbf{e}_2 + r_3 \mathbf{e}_3 - \eta_{\alpha}\mathbf{b}_0/2 + \eta_{\alpha}\mathbf{b}_{\mu}/2  \notag\\
     &= \mathbf{r}_{\alpha} + \eta_{\alpha}\mathbf{b}_{\mu}/2, \label{eq:SIPC_SIDC}
\end{align}
where $\eta_\alpha=1 \text{ if } \alpha=A$ \text{or} $\eta_\alpha=-1 \text{ if } \alpha=B$, $\mathbf{b}_\mu$ are vectors connecting the center of a down-pointing tetrahedron to the centers of its nearest up-pointing tetrahedrons
\begin{subequations}
\begin{align}
    \mathbf{b}_0 &= -\frac{1}{4}(1,1,1)\\
    \mathbf{b}_1 &= \frac{1}{4}(-1,1,1)\\
    \mathbf{b}_2 &= \frac{1}{4}(1,-1,1)\\
    \mathbf{b}_3 &= \frac{1}{4}(1,1,-1), \label{eq:bmu}
\end{align}
\end{subequations}
and $\mathbf{e}_i$ are the lattice basis vectors
\begin{subequations}
\begin{align}
    \mathbf{e}_0 &= (0,0,0)\\
    \mathbf{e}_1 &= \frac{1}{2}(0,1,1)\\
    \mathbf{e}_2 &= \frac{1}{2}(1,0,1)\\
    \mathbf{e}_3 &= \frac{1}{2}(1,1,0).
\end{align}
\end{subequations}
We have introduced $\mathbf{e}_0=\mathbf{0}$ for convenience (see e.g. Eq.~\eqref{eq:plaquettealgebra}). 

Spins are defined in sublattice-dependent local frames whose basis vectors are given in table \ref{tab: Local basis}
\begin{table}[!ht]
\caption{\label{tab: Local basis}%
Local sublattice basis vectors
}
\begin{ruledtabular}
\begin{tabular}{ccccc}
$\mu$ & 0 & 1  & 2  & 3 \\
\hline
$\hat{z}_{\mu}$ & $\frac{1}{\sqrt{3}}\left(1,1,1\right)$ & $\frac{-1}{\sqrt{3}}\left(-1,1,1\right)$  & $\frac{-1}{\sqrt{3}}\left(1,-1,1\right)$  & $\frac{-1}{\sqrt{3}}\left(1,1,-1\right)$   \\[2mm]
$\hat{y}_{\mu}$ & $\frac{1}{\sqrt{2}}\left(0,-1,1\right)$  & $\frac{1}{\sqrt{2}}\left(0,1,-1\right)$  & $\frac{-1}{\sqrt{2}}\left(0,1,1\right)$ & $\frac{1}{\sqrt{2}}\left(0,1,1\right)$  \\[2mm]
$\hat{x}_{\mu}$ & $\frac{1}{\sqrt{6}}\left(-2,1,1\right)$ & $\frac{-1}{\sqrt{6}}\left(2,1,1\right)$  & $\frac{1}{\sqrt{6}}\left(2,1,-1\right)$  & $\frac{1}{\sqrt{6}}\left(2,-1,1\right)$    \\
\end{tabular}
\end{ruledtabular}
\end{table}

\section{Transformation Properties of the Parton Operators \label{sec:Appendix_DOice}}

In the case where $J_{yy}$ is dominant for an octupolar spin ice, the generators of the pyrochlore space group act on the parton construction as:
\begin{widetext}
\begin{subequations}
\begin{align}
T_i: & \left\{\frac{1}{2} \Phi_{\mathbf{r}_A}^{\dagger} e^{i A_{\mathbf{r}_A, \mathbf{r}_A}+\mathbf{b}_\mu} \Phi_{\mathbf{r}_A+\mathbf{b}_\mu}, \frac{1}{2} \Phi_{\mathbf{r}_A+\mathbf{b}_\mu}^{\dagger} e^{-i A_{\mathbf{r}_A, \mathbf{r}_A+\mathbf{b}_\mu}} \Phi_{\mathbf{r}_A}, E_{\mathbf{r}_A, \mathbf{r}_A+\mathbf{b}_\mu}\right\} \notag\\
& \mapsto\left\{\frac{1}{2} \Phi_{T_i\left(\mathbf{r}_A\right)}^{\dagger} e^{i A_{T_i\left(\mathbf{r}_A\right), T_i\left(\mathbf{r}_A+\mathbf{b}_\mu\right)}} \Phi_{T_i\left(\mathbf{r}_A+\mathbf{b}_\mu\right)}, \frac{1}{2} \Phi_{T_i\left(\mathbf{r}_A+\mathbf{b}_\mu\right)}^{\dagger} e^{-i A_{T_{\ell}\left(\mathbf{r}_A\right), T_{\ell}\left(\mathbf{r}_A+\mathbf{b}_\mu\right)}} \Phi_{T_i\left(\mathbf{r}_A\right)}, E_{T_i\left(\mathbf{r}_A\right), T_i\left(\mathbf{r}_A+\mathbf{b}_\mu\right)}\right\}\\
\bar{C}_6: &\left\{\frac{1}{2} \Phi_{\mathbf{r}_A}^{\dagger} e^{i A_{\mathbf{r}_A, \mathbf{r}_A}+\mathbf{b}_\mu} \Phi_{\mathbf{r}_A+\mathbf{b}_\mu}, \frac{1}{2} \Phi_{\mathbf{r}_A+\mathbf{b}_\mu}^{\dagger} e^{-i A_{\mathbf{r}_A}, \mathbf{r}_A+\mathbf{b}_\mu} \Phi_{\mathbf{r}_A}, E_{\mathbf{r}_A, \mathbf{r}_A+\mathbf{b}_\mu}\right\} \notag\\
& \mapsto\left\{\frac{1}{2} \Phi_{\bar{C}_6\left(\mathbf{r}_A\right)}^{\dagger} e^{i A_{\bar{C}_6\left(\mathbf{r}_A\right), \bar{C}_6\left(\mathbf{r}_A+\mathbf{b}_\mu\right)}} \Phi_{\bar{C}_6\left(\mathbf{r}_A+\mathbf{b}_\mu\right)}, \frac{1}{2} \Phi_{\bar{C}_6\left(\mathbf{r}_A+\mathbf{b}_\mu\right)}^{\dagger} e^{-i A_{\bar{C}_6\left(\mathbf{r}_A\right), \bar{C}_6\left(\mathbf{r}_A+\mathbf{b}_\mu\right)}} \Phi_{\bar{C}_6\left(\mathbf{r}_A\right)}, E_{\bar{C}_6\left(\mathbf{r}_A\right), \bar{C}_6\left(\mathbf{r}_A+\mathbf{b}_\mu\right)}\right\} \\
S: &\left\{\frac{1}{2} \Phi_{\mathbf{r}_A}^{\dagger} e^{i A_{\mathbf{r}_A, \mathbf{r}_A}+\mathbf{b}_\mu} \Phi_{\mathbf{r}_A+\mathbf{b}_\mu}, \frac{1}{2} \Phi_{\mathbf{r}_A+\mathbf{b}_\mu}^{\dagger} e^{-i A_{\mathbf{r}_A, \mathbf{r}_A}+\mathbf{b}_\mu} \Phi_{\mathbf{r}_A}, E_{\mathbf{r}_A, \mathbf{r}_A+\mathbf{b}_\mu}\right\} \notag\\
& \mapsto\left\{-\frac{1}{2} \Phi_{S\left(\mathbf{r}_A\right)}^{\dagger} e^{i A_{S\left(\mathbf{r}_A\right), S\left(\mathbf{r}_A+\mathbf{b}_\mu\right)}} \Phi_{S\left(\mathbf{r}_A+\mathbf{b}_\mu\right)},-\frac{1}{2} \Phi_{S\left(\mathbf{r}_A\right)}^{\dagger} e^{i A_{S\left(\mathbf{r}_A\right), S\left(\mathbf{r}_A+\mathbf{b}_\mu\right)}} \Phi_{S\left(\mathbf{r}_A+\mathbf{b}_\mu\right)},E_{S\left(\mathbf{r}_A\right), S\left(\mathbf{r}_A+\mathbf{b}_\mu\right)}\right\}.
\end{align}
\end{subequations} 
\end{widetext}
However, if $J_{xx}$ is dominant, for dipolar spin ice, $S^\pm=S^y\pm iS^z$ and the screw operation would instead map $S: S^\pm\rightarrow S^\mp$. Under GMFT formalism, this swaps the spinon creation and annihilation operators and, therefore, will result in different PSG equations since this operation flips $\bar{A}\rightarrow -\bar{A}$. We will see in Appendix~\ref{sec:dipolar_ice} that this will result in different PSG classes.

\section{Classification of PSG under Fields in Different Directions\label{sec:AppendixPSG}}

\subsection{Generalities}
As discussed in Sec.~\ref{sec:PSGClass}, mean-field Ans\"atze that are related via some $U(1)$ gauge transformation correspond to the same physical wave function. As such, mean-field Ans\"atze needs only to be invariant under $\mathcal{G}_\mathcal{O}\mathcal{O}$, where $\mathcal{O}$ is some symmetry operation and $\mathcal{G}_\mathcal{O}:|\Psi_{\mathbf{r}_\alpha}\rangle \rightarrow e^{i\phi_\mathcal{O}(\mathbf{r}_\alpha)}|\Psi_{\mathbf{r}_\alpha}\rangle$ is some $U(1)$ transformation associated with the symmetry operation. 

The projective representations of the space group generators $\mathcal{G}_\mathcal{O}\mathcal{O}$ must follow the same underlying algebraic relation. The PSG classification of all fully symmetric states is therefore found by imposing such algebraic constraints. However, this is not enough to unambiguously find a representation of the fully symmetric states due to the fact that there is a gauge freedom in $\mathcal{G}_\mathcal{O}$. A PSG element $\mathcal{G}_\mathcal{O} \mathcal{O}\in \text{PSG}$ transforms under some general gauge transformation $\mathcal{G}:\ket{\Psi_{\mathbf{r}_\alpha}}=e^{i\phi(\mathbf{r}_\alpha)}\ket{\Psi_{\mathbf{r}_\alpha}}$ as
\begin{align}
   \mathcal{G}_\mathcal{O} \mathcal{O} \rightarrow \mathcal{G}\mathcal{G}_\mathcal{O} \mathcal{O} \mathcal{G}^{-1} &= \mathcal{G} \mathcal{G}_\mathcal{O} \mathcal{O} \mathcal{G}^{-1} \mathcal{O}^{-1}\mathcal{O} \\
   &= \mathcal{G}\mathcal{G}_\mathcal{O} \mathcal{G}^{-1}[\mathcal{O}^{-1}(\mathbf{r}_\alpha)]\mathcal{O},
\end{align}
where we used the relation
\begin{equation}
    \mathcal{O} \mathcal{G}\mathcal{O}^{-1} = \mathcal{G}[\mathcal{O}^{-1}(\mathbf{r}_\alpha)].
\end{equation}
Here $\mathcal{G}[\mathcal{O}^{-1}(\mathbf{r}_\alpha)]:\ket{\Psi_{\mathbf{r}_\alpha}}\rightarrow e^{i\phi(\mathcal{O}^{-1}(\mathbf{r}_\alpha))}\ket{\Psi_{\mathbf{r}_\alpha}}$
As such, the PSG phase transforms as
\begin{equation}
    \phi_\mathcal{O}(\mathbf{r}_\alpha) \rightarrow \phi_\mathcal{O}(\mathbf{r}_\alpha) + \phi(\mathbf{r}_\alpha) - \phi(\mathcal{O}^{-1}(\mathbf{r}_\alpha)).
\end{equation}
To obtain an unambiguous representation of the PSG class, we must fix the gauge freedoms. If we assume spatially isotropic and translationally invariant phase factors, there are 6 distinct gauge transformations corresponding to 2 diamond sublattices and 3 directions.
\begin{equation}
    \phi_{i,\beta}(\mathbf{r}_\alpha) =  \psi_{i,\beta} r_i \delta_{\alpha,\beta} \label{eq:gauge1},
\end{equation}
where $\psi_{i,\beta}$ is some $U(1)$ element. Furthermore, one can apply a sublattice-dependent gauge transformation:
\begin{equation}
    \bar{\phi}(\mathbf{r}_\alpha) =  \bar{\psi}_{\beta} \delta_{\alpha,\beta} \label{eq:gauge2},
\end{equation}
$\psi_{i,\beta}$, $\bar{\psi}_\beta \in [0, 2\pi)$. In total, we have 8 gauge degrees of freedom that we will use in the following sections to completely fix the phase factors along with site-independent $U(1)$ gauge transformation associated with each operation.

\subsection{\texorpdfstring{$\mathbf{B}\parallel[110]$}{\texttwoinferior}} \label{sec:AppendixPSG110}
\subsubsection{Algebraic Constraints}
Under the [110] field, the Zeeman couplings on the four pyrochlore sites are $\sqrt{\frac{2}{3}}h(1,0,0,-1)$. The point group of the pyrochlore lattice can be considered the direct product of the sublattice permutation group $S_4$ and the transformation group generated by inversion (i.e., $O_h\simeq S_4 \ltimes \mathbb{Z}_2$). As shown in Appendix~\ref{sec:Appendix_DOice}, screw operator $S$ will map $S^\pm\rightarrow-S^\pm$. In doing so, it exchanges two pyrochlore sites. We can think of this as an odd permutation in the $S_4$ group. On the other hand, the generator $\bar{C}_6$ is an even permutation under $S_4$ since it is a three-cycle (i.e. (123)=(13)(12)). As such, all odd permutations will map $S^\pm\rightarrow -S^\pm$, whereas even permutations will not. This results in rather unexpected symmetry operations under finite magnetic fields. For example, under a [110] field, naively speaking, one would expect a mirror symmetry that swaps the 1$^\mathrm{st}$ and the 2$^\mathrm{nd}$ pyrochlore sites $\sigma_{12}=\bar{C}_6^4S\bar{C}_6(S\bar{C}_6^{-1})^2$, since both Zeeman couplings are 0. However, this operation is odd and maps $S^\pm\rightarrow -S^\pm$. As such, $\sigma_{12}:\sqrt{\frac{2}{3}}h(1,0,0,-1)\rightarrow\sqrt{\frac{2}{3}}h(-1,0,0,1)$ and therefore is not a symmetry. 

The remaining symmetries under a [110] field are generated by Inversion $I=\bar{C}_6^3$ and a mirror symmetry $\sigma=S\bar{C}_6^3$, which swaps sites 0 and 3, along with the lattice translation symmetries $T_i$. The algebraic constraints for the generators are:

\begin{subequations}
\begin{align}
T_i T_{i+1} T_i^{-1} T_{i+1}^{-1}&=1, i\in \{1,2,3\}\\
T_i\sigma T_i^{-1} T_3 \sigma &= 1, i\in \{1,2\}\\
T_3\sigma T_3^{-1} \sigma &= 1\\
(IT_i)^2&=1, i\in \{1,2,3\}\\
T_3(I\sigma)^2&=1\\
\sigma^2 &= 1\\
I^2 &= 1
\end{align}
\end{subequations}

We can enrich these algebraic constraints by allowing a gauge transformation associated with each symmetry operation (i.e., $\mathcal{O}\rightarrow\mathcal{G}_\mathcal{O}\mathcal{O}$). Likewise, the unity of these algebraic constraints is also promoted to a $U(1)$ element.

\begin{widetext}
\begin{subequations} \label{eq:110PSGend}
\begin{align}
\mathcal{G}_{T_i} \mathcal{G}_{T_{i+1}}\left[T^{-1}\left(\mathbf{r}_A\right)\right] \mathcal{G}_{T_i}^{-1}\left[T_{i+1}^{-1}\left(\mathbf{r}_A\right)\right] \mathcal{G}_{T_{i+1}}^{-1}&=e^{i\psi_{T_i}} \\
\mathcal{G}_{T_i}\mathcal{G}_\sigma[T_i^{-1}(\mathbf{r}_\alpha)] \mathcal{G}_{T_i}^{-1}[\sigma T_3^{-1}(\mathbf{r}_\alpha)] \mathcal{G}_{T_3}[T_3\sigma(\mathbf{r}_A)] \mathcal{G}_\sigma^{-1}&=e^{i\psi_{\sigma T_i}} , i\in \{1,2\} \label{eq:sigma1}\\
\mathcal{G}_{T_3} \mathcal{G}_\sigma[T_3^{-1}(\mathbf{r}_\alpha)] \mathcal{G}_{T_3}[T_3\sigma(\mathbf{r}_A)] \mathcal{G}_\sigma^{-1}&=e^{i\psi_{\sigma T_3}} \label{eq:sigma2}\\
\mathcal{G}_I \mathcal{G}_{T_i}\left[\left(T_i I\right)^{-1}\left(\mathbf{r}_\alpha\right)\right] \mathcal{G}_I^{-1}\left[T_i^{-1}\left(\mathbf{r}_\alpha\right)\right] \mathcal{G}_{T_i}^{-1}&=e^{i\psi_{I T_i}}, i \in \{1, 2, 3\} \label{eq:inv1}
\\
\mathcal{G}_{T_3}\mathcal{G}_I[T_3^{-1}(\mathbf{r}_\alpha)]\mathcal{G}_\sigma[IT_3^{-1}(\mathbf{r}_\alpha)] \mathcal{G}_I[I\sigma(\mathbf{r}_\alpha)]\mathcal{G}_\sigma(\sigma(\mathbf{r}_\alpha))&=e^{i\psi_{I\sigma}} \\
\mathcal{G}_\sigma \mathcal{G}_\sigma[\sigma(\mathbf{r}_\alpha)] &=e^{i\psi_{\sigma}}\label{eq:110Isigma}\\
\mathcal{G}_I \mathcal{G}_I[I(\mathbf{r}_\alpha)] &=e^{i\psi_{I}} \label{eq:inv2}.
\end{align}
\end{subequations} 
\end{widetext}
All of the $\psi$ parameters above are in $[0,2\pi)$.

\subsubsection{Solution to the PSG}
A common gauge fixing is to first fix the phase factor associated with translation symmetries. $\phi_{T_1}(\mathbf{r}_1,\mathbf{r}_2,\mathbf{r}_3)=\phi_{T_2}(0,\mathbf{r}_2,\mathbf{r}_3)=\phi_{T_i}(0,0,\mathbf{r}_3)=0$ by fixing $\psi_{i,\alpha}$ in equation~\eqref{eq:gauge1}. The associated PSG solution for the lattice translation is therefore
\begin{subequations}\label{eq:PSGT}
\begin{align}
\phi_{T_1}\left(\mathbf{r}_\alpha\right) & =0 \\
\phi_{T_2}\left(\mathbf{r}_\alpha\right) & =-\psi_{T_1} r_1 \\
\phi_{T_3}\left(\mathbf{r}_\alpha\right) & =\psi_{T_3} r_1-\psi_{T_2} r_2.
\end{align}
\end{subequations}
For the purpose of this study, since magnetic fields never break inversion symmetry, let us consider inversion symmetry first by solving Eq.~\eqref{eq:inv1}. This equation imposes the relation
\begin{equation}
    \phi_I(\mathbf{r}_\alpha) = \psi_{IT_1} r_1 + \psi_{IT_2} r_2 + \psi_{IT_3} r_3 + \phi_I(\mathbf{0}_\alpha)
\end{equation}
and the requirement that $\psi_{T_1}=n_1\pi$,  $\psi_{T_2}=n_2\pi$,  $\psi_{T_3}=n_3\pi$, $n_1, n_2, n_3 \in \mathbb{Z}_2$. Eq~\eqref{eq:inv2} then yields
\begin{equation}
    \phi_I(\mathbf{0}_A) + \phi_I(\mathbf{0}_B) = \psi_I.\label{eq:intra1}
\end{equation}
Similarly, we solve for $\phi_\sigma$ via Eqs.~\eqref{eq:sigma1} and \eqref{eq:sigma2} to obtain
\begin{align}
    \phi_\sigma(\mathbf{r}_\alpha)&=n_2\pi(r_1+r_2)(r_1+r_2+1)/2 \\
    &- \psi_{\sigma T_1} r_1 - \psi_{\sigma T_2} r_2 + \psi_{\sigma T_3} r_3 + \phi_\sigma(\mathbf{0}_\alpha).
\end{align}
and that $n_2=n_3$. The remaining constraints yield
\begin{subequations}
\begin{align}
&\psi_{\sigma T_3} = 2\psi_{\sigma T_1} = 2\psi_{\sigma T_2}\\
&2\phi_\sigma(\mathbf{0}_A)= \psi_\sigma \label{eq:intra2}\\
&2\phi_\sigma(\mathbf{0}_B)-\psi_{\sigma T_3}= \psi_\sigma\label{eq:intra3}\\
&\psi_{IT_3}+\psi_{\sigma T_3} = 0\\
&\phi_I(\mathbf{0}_A)+ \phi_I(\mathbf{0}_B)+\phi_\sigma(\mathbf{0}_A)+\phi_\sigma(\mathbf{0}_B)\notag\\
&\qquad\qquad\qquad\qquad\qquad-\psi_{IT_3}-\psi_{\sigma T_3}=\psi_{\sigma I}
\end{align}
\end{subequations}
By applying site-independent gauge transformation, we can fix $\psi_{\sigma T_1}=0$, which implies that $\psi_{IT_3}=\psi_{\sigma T_3}=\psi_{\sigma T_2}=0$ and $\phi_\sigma{\mathbf{0}_A}=\phi_\sigma(\mathbf{0}_B)$. We can then make a sublattice-dependent gauge transformation such that $\phi_\sigma(\mathbf{0}_A)=0=\psi_\sigma=\phi_\sigma(\mathbf{0}_B)$, which implies that $\psi_{\sigma I}=\psi_{I}$. We can finally make a gauge transformation such that $\phi_I(\mathbf{0}_A)=0$. Therefore, the final PSG solution is
\begin{subequations}
\begin{align}
&\phi_{T_1}(\mathbf{r}_\alpha) = 0\\
&\phi_{T_2}(\mathbf{r}_\alpha) = n_1\pi r_1\\
&\phi_{T_3}(\mathbf{r}_\alpha) = n_2\pi (r_1+r_2)\\
&\phi_\sigma(\mathbf{r}_\alpha) = n_2\pi (r_1+r_2)(r_1+r_2+1)/2\\
&\phi_I(\mathbf{r}_\alpha) = \psi_{IT_1}r_1 + \psi_{IT_2}r_2 + \psi_I \delta_{\alpha, B}.
\end{align}
\end{subequations}

\subsection{\texorpdfstring{$\mathbf{B}\parallel[111]$}{\texttwoinferior}}\label{sec:AppendixPSG111}

\subsubsection{Algebraic Constraints}

The only generator for remaining symmetry under a [111] field (besides lattice translations) is $\bar{C}_6$. The associated algebraic constraints (besides the ones from translations symmetry that have already been considered in the previous section) are
\begin{subequations}
\begin{align}
\bar{C}_6 T_i \bar{C}_6^{-1} T_{i+1}&=1 \;\forall i\in{1,2,3}\\
\bar{C}_6 ^6&=1.
\end{align}
\end{subequations}
The corresponding gauge-enriched equations are
\begin{subequations}
\begin{align}
\left(\mathcal{G}_{\bar{C}_6} \bar{C}_6\right)\left(\mathcal{G}_{T_i} T_i\right)\left(\mathcal{G}_{\bar{C}_6} \bar{C}_6\right)^{-1}\left(\mathcal{G}_{T_{i+1}} T_{i+1}\right) &= e^{i\psi_{CT_i}}\label{eq:CT}\\
\left(\mathcal{G}_{\bar{C}_6} \bar{C}_6\right)^6 &= e^{i\psi_{C_6}}. \label{eq:C6}
\end{align}
\end{subequations}  

\subsubsection{Solution to the PSG}

The lattice translation constraints yield the same thing as above
\begin{subequations}\label{eq:PSGwithInv}
\begin{align}
\phi_{T_1}\left(\mathbf{r}_\alpha\right) & =0 \\
\phi_{T_2}\left(\mathbf{r}_\alpha\right) & =-\psi_{T_1} r_1 \\
\phi_{T_3}\left(\mathbf{r}_\alpha\right) & =\psi_{T_3} r_1-\psi_{T_2} r_2.
\end{align}
\end{subequations}
Solving equation~\eqref{eq:CT} yields
\begin{equation}
\begin{aligned}
\phi_{\bar{C}_6}\left(\mathbf{r}_\alpha\right)&=\phi_{\bar{C}_6}\left(\mathbf{0}_\alpha\right)-r_2 \psi_{\bar{C}_6 T_1}-r_3 \psi_{\bar{C}_6 T_2}-r_1 \psi_{\bar{C}_6 T_3}\\&-\psi_{T_1}\left(r_1 r_2-r_1 r_3\right)
\end{aligned}
\end{equation}
with the constraint that $\psi_{T_1}=\psi_{T_2}=\psi_{T_3}$. Furthermore, from solving \eqref{eq:C6}, we obtain 
\begin{align}
    3\phi_{C_3}\left(\mathbf{0}_A\right)+3\phi_{C_3}\left(\mathbf{0}_B\right)&=\psi_{C_6}.
\end{align}
and $\psi_{T_1}=n_1\pi$ where $n_1\in\{0,1\}$.
Again, we can use gauge transformations to fix $\phi_{C_3}(\mathbf{0}_A)=0$, which is also possible via the sublattice gauge transformation. We arrive at the PSG solution
\begin{subequations}
\begin{align}
&\phi_{T_1}(\Bar{r}_\alpha) = 0\\
&\phi_{T_2}(\Bar{r}_\alpha) = n_1\pi r_1\\
&\phi_{T_3}(\Bar{r}_\alpha) = n_1\pi (r_1+r_2)\\
&\phi_{C}(\Bar{r}_\alpha) = n_1\pi r_1(r_2+r_3) + \frac{1}{3}\psi_{C_6}\delta_{\alpha,B}.
\end{align}
\end{subequations}

\subsection{\texorpdfstring{$\mathbf{B}\parallel[001]$}{\texttwoinferior}}\label{sec:AppendixPSG001}

\subsubsection{Algebraic Constraints}
Here, the generator of the remaining symmetries beyond $T_i$ and $I$ is a $\bar{C}_4$ improper rotation along the $[\frac{1}{2}\frac{1}{2}0]$ direction followed by a mirror reflection along the $(x,y,1/2)$ plane, $\bar{C}_4=\bar{C}_6^2S\bar{C}_6^{-1}$. The additional algebraic constraints, besides the ones for translation and inversion, are
\begin{subequations}
\begin{align}
\bar{C}_4^4&=1\\
T_2^{-1}\bar{C}_4 T_{1}^{-1} \bar{C}_4^{-1}&=1\\
T_{3} T_2^{-1}\bar{C}_4 T_{2}^{-1} \bar{C}_4^{-1}&=1\\
T_{1} T_2^{-1}\bar{C}_4 T_{3}^{-1} \bar{C}_4^{-1}&=1\\
T_2 I \bar{C}_4 I \bar{C}_4^{-1}&=1\\
T_3^{-1} (\bar{C}_4^2I)^2&=1.
\end{align}
\end{subequations}
The associated gauge-enriched equations are
\begin{widetext}
\begin{subequations}
\begin{align}
\mathcal{G}_{\bar{C}_4} \mathcal{G}_{\bar{C}_4}\left[\bar{C}_4^{-1}(\mathbf{r}_\alpha)\right] \mathcal{G}_{\bar{C}_4}\left[\bar{C}_4^2(\mathbf{r}_\alpha)\right]\mathcal{G}_{\bar{C}_4}\left[\bar{C}_4(\mathbf{r}_\alpha)\right] &= e^{i\psi_{\bar{C}_4}} \\
\mathcal{G}_{T_2}^{-1}\left[T_2(\mathbf{r}_\alpha)\right] \mathcal{G}_{\bar{C}_4}\left[T_2(\mathbf{r}_\alpha)\right]\mathcal{G}_{T_1}^{-1}\left[\bar{C}_4^{-1}(\mathbf{r}_\alpha)\right] \mathcal{G}_{\bar{C}_4}^{-1} &= e^{i\psi_{\bar{C}_4T_2}} \label{eq:ST2}\\
\mathcal{G}_{T_3} \mathcal{G}_{T_2}^{-1}\left[T_2 T_3^{-1}(\mathbf{r}_\alpha)\right] \mathcal{G}_{\bar{C}_4}\left[T_2 T_3^{-1}(\mathbf{r}_\alpha)\right] \mathcal{G}_{T_2}^{-1}\left[\bar{C}_4^{-1}(\mathbf{r}_\alpha)\right] \mathcal{G}_{\bar{C}_4}^{-1} &= e^{i\psi_{\bar{C}_4 T_3}}\label{eq:ST3}\\
\mathcal{G}_{T_1} \mathcal{G}_{T_2}^{-1}\left[T_2 T_1^{-1}(\mathbf{r}_\alpha)\right]\mathcal{G}_{\bar{C}_4}\left[T_2 T_1^{-1}\right]\mathcal{G}_{T_3}^{-1}\left[\bar{C}_4^{-1}(\mathbf{r}_\alpha)\right] \mathcal{G}_{\bar{C}_4}^{-1} &= e^{i\psi_{\bar{C}_4 T_1}}\label{eq:ST1}\\
\mathcal{G}_{T_2} \mathcal{G}_I\left[T_2^{-1}(\mathbf{r}_\alpha)\right] \mathcal{G}_{\bar{C}_4}\left[I T_2^{-1}(\mathbf{r}_\alpha)\right] \mathcal{G}_I\left[I \bar{C}_4^{-1}(\mathbf{r}_\alpha)\right] \mathcal{G}_{\bar{C}_4}^{-1}&= e^{i\psi_{\bar{C}_4I}}\\
\mathcal{G}_{T_3}^{-1}[T_3(\mathbf{r}_\alpha)]\mathcal{G}_{\bar{C}_4}[T_3(\mathbf{r}_\alpha)]\mathcal{G}_{\bar{C}_4}[\bar{C}_4^{-1}T_3(\mathbf{r}_\alpha)] \mathcal{G}_I[T_3\bar{C}_4^{-2}(\mathbf{r}_\alpha)]\mathcal{G}_{\bar{C}_4}[\bar{C}_4^2I(\mathbf{r}_\alpha)]\mathcal{G}_{\bar{C}_4}[\bar{C}_4I(\mathbf{r}_\alpha)]\mathcal{G}_I[I(\mathbf{r}_\alpha)]&=e^{i\psi_{\bar{C}_4 IT}}.
\end{align}
\end{subequations}    
\end{widetext}

\subsubsection{Solution to the PSG}
We can start from the PSG solution from the symmetry group generated by translation an inversion as in Eq.~\eqref{eq:PSGwithInv}. Solving Eqs. \eqref{eq:ST2}, \eqref{eq:ST3}, and \eqref{eq:ST1} gives

\begin{align}
&\psi_{\bar{C}_4}(\mathbf{r}_\alpha) = \frac{n_1\pi}{2}(r_1(1+r_1)+r_3(1+r_3))^2 +n_1\pi r_1 r_2\notag\\
&+(\psi_{\bar{C}_4T_2}-\psi_{\bar{C}_4T_1}r_1)+\psi_{\bar{C}_4T_2}r_2+(\psi_{\bar{C}_4T_1}-\psi_{\bar{C}_4T_3}r_3)\notag\\
&+n_1\pi(r_1+r_3)\delta_{\alpha,B}+\psi_{\bar{C}_4}(\mathbf{0}_\alpha),
\end{align}
and $n_1=n_2=n_3$.

The algebraic constraints on the order of the improper rotation gives
\begin{subequations}
\begin{align}
4\phi_{\bar{C}_4}(\mathbf{0}_B) -\psi_{\bar{C}_4T_1}+ \psi_{\bar{C}_4T_2} - \psi_{\bar{C}_4T_3}&= \psi_{\bar{C}_4}\\
4\phi_{\bar{C}_4}(\mathbf{0}_A) &= \psi_{\bar{C}_4}.
\end{align}
\end{subequations}
and the algebraic constraints involving the inversion operation yield
\begin{widetext}
\begin{subequations}
\begin{align}\label{eq:A001constraints}
 \psi_{I T_1}+\psi_{I T_2}-2 \psi_{\bar{C}_4 T_2}&=0 \\
 2 \psi_{I T_1}-\psi_{I T_3}+2 \psi_{\bar{C}_4 T_1}-2 \psi_{\bar{C}_4 T_2}&=0 \\
 \psi_{I T_1}-\psi_{I T_2}+\psi_{I T_3}-2 \psi_{\bar{C}_4 T_2}+2 \psi_{\bar{C}_4 T_3}&=0 \\
 \psi_{I T_1}-\psi_{I T_2}-\psi_{I T_3}+2 \psi_{\bar{C}_4 T_1}&=0 \\
-2 \psi_{I T_3}+2 \psi_{\bar{C}_4 T_1}+2 \psi_{\bar{C}_4 T_2}-2 \psi_{\bar{C}_4 T_3}&=0\\
 \phi_I\left(\mathbf{0}_A\right)+\phi_I\left(\mathbf{0}_B\right)-\phi_{\bar{C}_4 }\left(\mathbf{0}_A\right)+\phi_{\bar{C}_4 }\left(\mathbf{0}_B\right)-\psi_{I T_2}+\psi_{\bar{C}_4 T_2}&=\psi_{\bar{C}_4  I} \\
\phi_I\left(\mathbf{0}_A\right)+\phi_I\left(\mathbf{0}_B\right)+\phi_{\bar{C}_4}\left(\mathbf{0}_A\right)-\phi_{\bar{C}_4}\left(\mathbf{0}_B\right)-\psi_{I T_1}-\psi_{I T_2}+\psi_{\bar{C}_4 T_2}&=\psi_{\bar{C}_4 I} \\
\phi_I\left(\mathbf{0}_A\right)+\phi_I\left(\mathbf{0}_B\right)+2\left(\phi_{\bar{C}_4}\left(\mathbf{0}_A\right)+\phi_{\bar{C}_4}\left(\mathbf{0}_B\right)\right)+\psi_{I T_3}+\psi_{\bar{C}_4 T_1}+3 \psi_{\bar{C}_4 T_2}-2 \psi_{\bar{C}_4 T_3}&=\psi_{\bar{C}_4 IT} \\
\phi_I\left(\mathbf{0}_A\right)+\phi_I\left(\mathbf{0}_B\right)+2\left(\phi_S\left(\mathbf{0}_A\right)+\phi_S\left(\mathbf{0}_B\right)\right)+2\left(\psi_{I T_3}+\psi_{\bar{C}_4 T_2}\right)-\psi_{\bar{C}_4 T_3}&=\psi_{\bar{C}_4 IT}
\end{align}
\end{subequations}    
\end{widetext}

Let us fix the gauge such that $\phi_{\bar{C}_4}(\mathbf{0}_A)=\phi_{\bar{C}_4}(\mathbf{0}_B)$. This implies that $\psi_{IT_1}=0$. Then by fixing the gauge such that $\psi_{\bar{C}_4T_1}=\psi_{\bar{C}_4T_3}=0$, we enforce that $\psi_{\bar{C}_4T_2}=\psi_{IT_1}=\psi_{IT_3}=0$. Finally, by fixing $\psi_{\bar{C}_4 I}=0$ and $\phi_I(\mathbf{0}_A)=0$, we get that $\phi_I(\mathbf{0}_A)=\phi_I(\mathbf{0}_B)=0$. 

Finally, we arrive at the PSG solution:
\begin{subequations}
\begin{align}
&\phi_{T_1}(\mathbf{}{r}_\alpha) = 0\\
&\phi_{T_2}(\mathbf{r}_\alpha) = n_1\pi r_1\\
&\phi_{T_3}(\mathbf{r}_\alpha) = n_1\pi (r_1+r_2)\\
&\phi_{\bar{C}_4}(\mathbf{r}_\alpha) = \frac{n_1\pi}{2}(-r_1(1+r_1)+r_3(1+r_3))\notag \\
\quad\quad&+n_1\pi r_1 r_2+n_1\pi(r_1+r_3)\delta_{\alpha,B}+\psi_{\bar{C}_4}/4\\
&\phi_I(\mathbf{r}_\alpha) = 0
\end{align}
\end{subequations}

\subsection{Mean Gauge Configuration}
After getting all the PSG solutions, we can now find the fully symmetric mean-field solution by looking at how they are related under the remaining symmetry operations. Indeed, the MF Hamiltonian has to be invariant under the projective transformations $\mathcal{G}_\mathcal{O}\mathcal{O}$. This determines the 
corresponding mean-field Ans\"atze $\mathcal{G}_{\bar{A}}(\mathbf{r}_A,\mathbf{r}_A+b_\mu)= e^{iA_{\mathbf{r}_A,\mathbf{r}_A+b_\mu}}$. More specifically, invariance of the MF Hamiltonian under $\mathcal{G}_\mathcal{O}\mathcal{O}$ implies
\begin{equation}
\begin{aligned}
    \mathcal{G}_{\bar{A}} &(O(\mathbf{r}_A), O(\mathbf{r}_A + \mathbf{b}_\mu)) \\
    &= \mathcal{G}_O^\dagger(O(\mathbf{r}_A)) \mathcal{G}_{\bar{A}} (\mathbf{r}_A, \mathbf{r}_A + \mathbf{b}_\mu) \mathcal{G}_O(O(\mathbf{r}_A)).\label{eq:gaugeconfigrelateion}
\end{aligned}
\end{equation}
From these requirements, the gauge field configuration on the entire lattice can be determined for a given PSG class. To do so, the value of the gauge field background is arbitrarily fixed on representative bonds that are not related to each other but are related by all other bonds on the lattice by symmetry operations. In the presence of translation symmetry, this inequivalent set can be taken to contain at most the for bonds of the unit cell at the origin $\bar{A}_{\mathbf{0}_A, \mathbf{0}_A+\mathbf{b}_\mu}:=\bar{A}_\mu$. In many cases, these bonds might be related by symmetry operations. 

For the [110] field, the bonds $\bar{A}_{\mathbf{0}_A, \mathbf{0}_A+\mathbf{b}_0}$, $\bar{A}_{\mathbf{0}_A, \mathbf{0}_A+\mathbf{b}_1}$, and $\bar{A}_{\mathbf{0}_A, \mathbf{0}_A+\mathbf{b}_2}$ cannot be related by the remaining symmetry transformations. The mean-field configuration for a specific PSG class is then specified by three parameters $\bar{A}_0$, $\bar{A}_1$, and $\bar{A}_2$:
\begin{subequations}\label{eq:A110}
\begin{align}
& A\left[\left(r_1, r_2, r_3\right)_A,\left(r_1, r_2, r_3\right)_B\right]=\bar{A}_0  \\
& A\left[\left(r_1, r_2, r_3\right)_A,\left(r_1+1, r_2, r_3\right)_B\right]=\bar{A}_1+n_1 \pi r_2+n_2\pi r_3   \\
& A\left[\left(r_1, r_2, r_3\right)_A,\left(r_1, r_2+1, r_3\right)_B\right]=\bar{A}_2+n_2 \pi r_3  \\
& A\left[\left(r_1, r_2, r_3\right)_A,\left(r_1, r_2, r_3+1\right)_B\right]=\bar{A}_0.
\end{align}
\end{subequations}

Similarly, we can only mix the bonds $\bar{A}_{\mathbf{0}_A, \mathbf{0}_A+\mathbf{b}_1}$, $\bar{A}_{\mathbf{0}_A, \mathbf{0}_A+\mathbf{b}_2}$ and $\bar{A}_{\mathbf{0}_A, \mathbf{0}_A+\mathbf{b}_3}$ by symmetry operations in the presence of a [111] field. The gauge field background configuration of the associated PSG class then depends on two parameters $\bar{A}_0$ and $\bar{A}_1$:
\begin{subequations}\label{eq:A111}
\begin{align}
& A\left[\left(r_1, r_2, r_3\right)_A,\left(r_1, r_2, r_3\right)_B\right]=\bar{A}_0 \\
& A\left[\left(r_1, r_2, r_3\right)_A,\left(r_1+1, r_2, r_3\right)_B\right]=\bar{A}_1+n_1 \pi\left(r_2+r_3\right) \\
& A\left[\left(r_1, r_2, r_3\right)_A,\left(r_1, r_2+1, r_3\right)_B\right]=\bar{A}_1+n_1 \pi r_3 \\
& A\left[\left(r_1, r_2, r_3\right)_A,\left(r_1, r_2, r_3+1\right)_B\right]=\bar{A}_1
\end{align}
\end{subequations}

Finally, for the $[001]$ field, $\bar{C}_4$ can indeed relate all 4 sites. Therefore, the gauge field configuration would depend on the parameter $\bar{A}_{0}$:
\begin{subequations}\label{eq:A001}
\begin{align}
& A\left[\left(r_1, r_2, r_3\right)_A,\left(r_1, r_2, r_3\right)_B\right]=\bar{A}_0 \\
& A\left[\left(r_1, r_2, r_3\right)_A,\left(r_1+1, r_2, r_3\right)_B\right]=\bar{A}_0+n_1 \pi (r_2+r_3) \\
& A\left[\left(r_1, r_2, r_3\right)_A,\left(r_1, r_2+1, r_3\right)_B\right]=\bar{A}_0+ n_1 \pi r_3 \\
& A\left[\left(r_1, r_2, r_3\right)_A,\left(r_1, r_2, r_3+1\right)_B\right]=\bar{A}_0
\end{align}
\end{subequations}

As argued in the next section, the gauge-invariant fluxes (as well as any physical observable) do not ultimately depend on the above parameters $\overline{A}_\mu$. Since they do not contribute anything to the underlying physics, they can simply be fixed to zero in all PSG classes.  

\subsection{Minimal Constraints on Plaquette Fluxes\label{sec:AppendixPSG0}}

As mentioned in the main text, the fluxes on the hexagonal plaquettes must be translational invariant and $\pi$-multiple in the presence of translation and inversion symmetries. One can see the reasons by working our way back as we solve the PSG algebraic constraints. If we only impose lattice translation symmetry, we arrive at the PSG solution in Eq.~\eqref{eq:PSGT}. If we substitute this solution into Eq.~\eqref{eq:plaquettealgebra}, we can see that the spatial dependence of the fluxes disappears. Namely, the plaquettes in the aforementioned ordered sequence take on some constant value $(-\psi_{T_1}, -\psi_{T_1}+\psi_{T_2}-\psi_{T_3}, \psi_{T_2}, \psi_{T_3})$. Now, if we additionally consider inversion symmetry, we find the requirement that $\psi_{T_1}=n_1\pi$, $\psi_{T_2}=n_2\pi$, $\psi_{T_3}=n_3\pi$. Therefore, the inequivalent fluxes are $(n_1\pi, n_1\pi+n_2\pi+n_3\pi, n_2\pi, n_3\pi)$. As such, the $\pi$ ``quantization" of the fluxes comes from PSG constraints from inversion symmetry. In fact, any symmetry operations that contain an inversion operator will result in this $\pi$ ``quantization" (for example, $\sigma=SI$). 

\subsection{PSG for Dipolar Spin Ice under Magnetic Fields\label{sec:dipolar_ice}}

In the case where $J_{xx}$ is dominant, the screw operation $S$ will instead map creation operators to annihilation operators and vice versa. Therefore, the PSG will differ for SG with generators containing $S$. These operators are $\sigma$ under $[110]$ fields and $\bar{C}_4$ under $[001]$ fields. Therefore, we expect different PSG results for these fields. After going through similar derivations, it turns out that under a $[110]$ field, the PSG solution is:
\begin{subequations}
\begin{align}
\phi_{T_1}(\mathbf{r}_\alpha) &= 0\\
\phi_{T_2}(\mathbf{r}_\alpha) &= n_1\pi r_1\\
\phi_{T_3}(\mathbf{r}_\alpha) &= n_2\pi (r_1+r_2)\\
\phi_\sigma(\mathbf{r}_\alpha) &= n_2\pi (r_1+r_2)(r_1+r_2+1)/2\notag\\
&\quad -\psi_{\sigma T_1}r_1-\psi_{\sigma T_2}r_2\label{eq:sxx1}\\
\phi_I(\mathbf{r}_\alpha) &= n_I\pi(r_1+r_2)+ \psi_I/2\label{eq:sxx2}.
\end{align}
\end{subequations}
where $n_I\in\{0,1\}$. The PSG solutions under a $[001]$ field is:
\begin{subequations}
\begin{align}
&\phi_{T_1}(\mathbf{}{r}_\alpha) = 0\\
&\phi_{T_2}(\mathbf{r}_\alpha) = n_1\pi r_1\\
&\phi_{T_3}(\mathbf{r}_\alpha) = n_1\pi (r_1+r_2)\\
&\phi_{\bar{C}_4}(\mathbf{r}_\alpha) = \frac{n_1\pi}{2}(-r_1(1+r_1)+r_3(1+r_3))\notag \\
&\qquad\quad+n_1\pi r_1 r_2+(n_1\pi(r_1+r_3))\delta_{\alpha,B}+\psi_{C}\label{eq:sxx3}\\
&\phi_I(\mathbf{r}_\alpha) = n_I\pi.\label{eq:sxx4}
\end{align}
\end{subequations}
where $n_I\in\mathbb{Z}_2$, $\psi_{C}\in U(1)$. As such, the PSG solution for the dipolar case differs from that of the octupolar case. 
Despite this difference, both result in the same mean-field gauge configuration $\bar{A}_{\mathbf{r}_\alpha}$ as in Eqs.~\eqref{eq:A110} and~\eqref{eq:A001} up to a gauge transformation. For the XXZ model, this will not amount to a difference in the fluxes of hexagonal plaquettes (i.e., the MF Ans\"atze are equivalent). However, when considering the XYZ model, other mean-field parameters beyond the gauge field background are present. These may differ in the octupolar and dipolar spin ice cases, thus resulting in distinct MF Hamiltonians in both cases. In particular, there may be more states in the dipolar case as labeled by the additional PSG indices $n_S$ and $\psi_{\bar{C}_4}$.

\section{Competition with the Gauge Background\label{sec:competition}}
 
The phase diagrams obtained with GMFT should be contrasted with the recent results of Ref.~\cite{yan2023experimentally}. In this study, Yan, Sanders, Castelnovo, and Nevidomskyy studied a similar model~\eqref{eq:H_XXZ} perturbatively by projecting interactions into the spin ice (i.e., chargeless) manifold. This method effectively removes (i.e., integrates out) the non-vanishing charge configuration and yields an effective Hamiltonian of the form
fields~\eqref{eq:H_XXZ}.
\begin{align}
    &\mathcal{H}_{\text{eff}} \sim \notag\\
    &\sum_{\mathbf{r}_A}\sum_{\mu\nu\lambda\sigma} \left(- \frac{12J_\pm^3}{J_{yy}^2} + \frac{5J_\pm^2}{J_{yy}^3} (\varepsilon_{\mu\nu\lambda\sigma}\hat{\mathbf{z}}_\sigma \cdot \mathbf{B})^2  \right) \cos(F_{\mu\nu\lambda}(\mathbf{r}_A))\label{eq:ringexchange}
\end{align} 
where $\epsilon_{\mu\nu\lambda\sigma}$ is the Levi-Civita symbol. We see that the leading term stemming from the Zeeman coupling always increases the effective ring exchange coupling and thereby favors the $\pi$-flux phase. From this perturbative treatment, one expects a field-induced phase transition from 0-flux to $\pi$-flux QSI, in contrast to the GMFT predictions. Furthermore, Ref.~\cite{yan2023experimentally} predicts a stable staggered flux phase with a different flux configuration under a $[110]$ field.

To understand the possible origin of these discrepancies, one should remember the different approximations in both approaches. GMFT captures the spinon dynamics but ignores contributions from the gauge bosons (i.e., the fluctuating gauge field sector). The energetics of different Ans\"atze in GMFT are thus determined by the spinon dispersion, which depends upon the static gauge field background. In contrast, the perturbative treatment of Ref.~\cite{yan2023experimentally} ignores defect tetrahedrons (i.e., the matter field) by restricting the theory to the spin ice manifold. The energetics are then fully determined by the gauge field sector. Therefore, the discrepancies highlighted above in the presence of a magnetic field seem to indicate the existence of some competition between the spinon and gauge field energy sectors. Which approach ultimately wins this energetic tug of war will need to be determined by future unbiased numerical investigations or theoretical approaches that incorporate both contributions. 
\begin{figure}
    \centering
    \includegraphics[width=\linewidth]{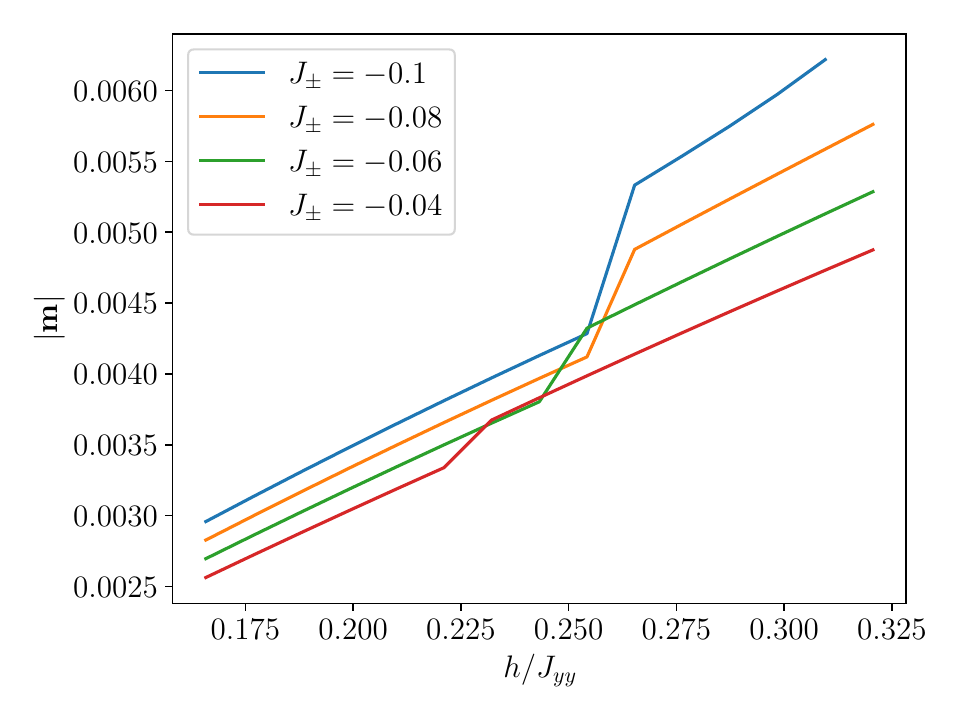}
    \caption{Normalized magnetization $g_{zz}\mu_B$ per pyrochlore site near from $J_\pm=-0.04$ to $J_\pm=-0.1$  under a [110] magnetic field. The abrupt jump shows a first-order phase transition from the $\pi$-flux phase to the $(0,\pi,\pi,0)$ phase.}
    \label{fig:magn}
\end{figure}

\begin{figure}
    \centering
    \includegraphics[width=\linewidth]{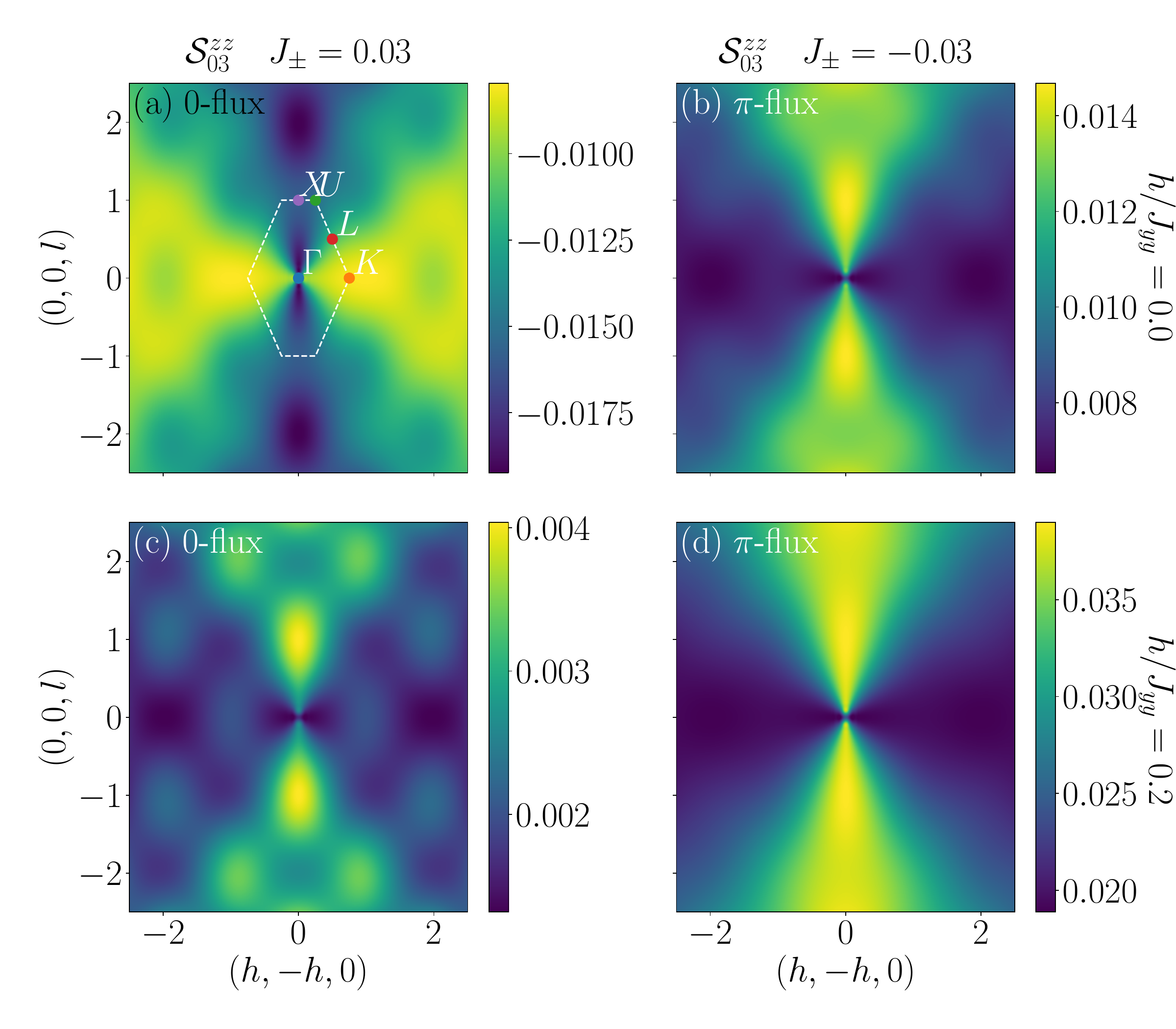}
    \caption{Sublattice SSSF in the global frame for intra-$\alpha$-chain correlation $\mathcal{S}^{zz}_{03}$ under a $[110]$ field when $J_\pm=0.03$ (a), (c) for $h=0J_{yy}$ and $h=0.2 J_{yy}$ respectively; when $J_\pm=-0.03$ (b), (d) for $h=0J_{yy}$ and $h=0.2 J_{yy}$ respectively. The First Brillouin zone is highlighted in white.}
    \label{fig:sublattice110_1}
\end{figure}

\begin{figure}
    \centering
    \includegraphics[width=\linewidth]{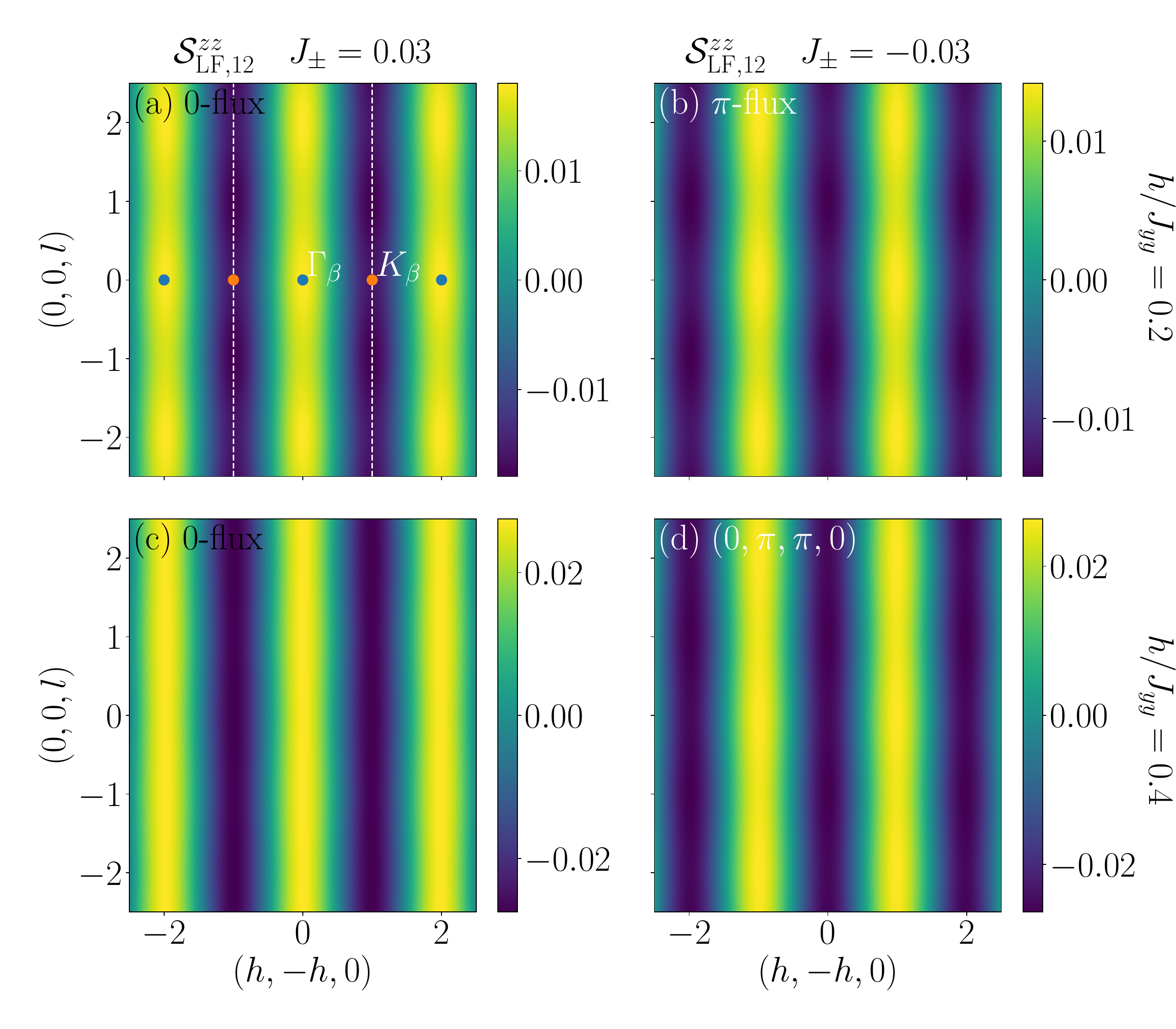}
    \caption{Sublattice SSSF in the local frame for intra-$\beta$-chain correlation $\mathcal{S}^{zz}_{\mathrm{LF},12}$ under a $[110]$ field when $J_\pm=0.03$ (a), (c) for $h=0.2J_{yy}$ and $h=0.4 J_{yy}$ respectively; when $J_\pm=-0.03$ (b), (d) for $h=0.2J_{yy}$ and $h=0.4 J_{yy}$ respectively. First Brillouin zone of the $\beta$ chain is highlighted in white. The $\beta$ chain $\Gamma$ points, $\Gamma_\beta$, are highlighted in blue and the $\beta$ chain $K$ points, $K_\beta$, are highlighted in orange.}
    \label{fig:sublattice110_3}
\end{figure}

\begin{figure}
    \centering
    \includegraphics[width=\linewidth]{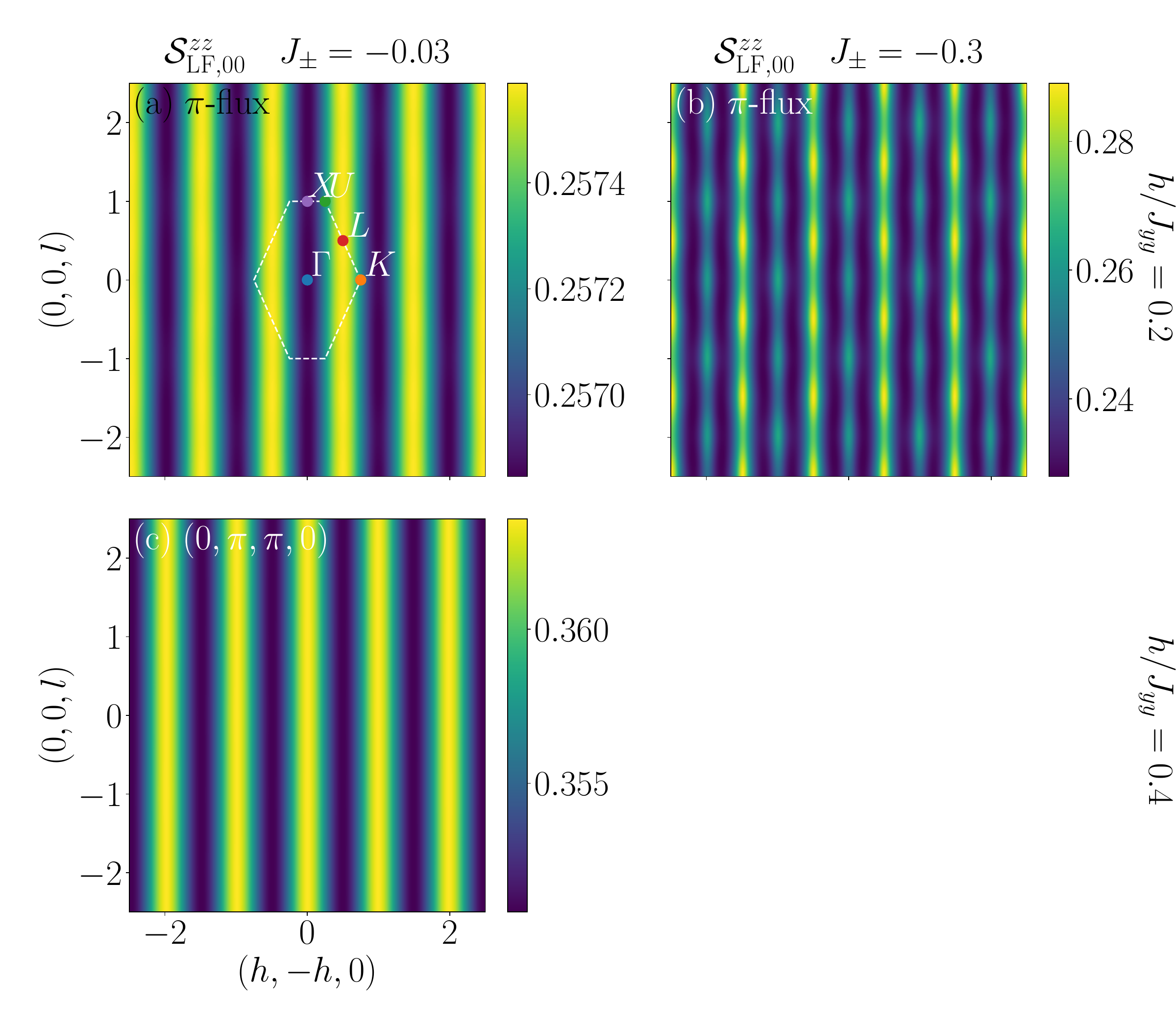}
    \caption{Sublattice SSSF in the local frame for inter-$\alpha$-chain correlation $\mathcal{S}^{zz}_{\mathrm{LF},00}$ under a $[110]$ field when $J_\pm=-003$ (a), (c) for $h=0.2 J_{yy}$ and $h=0.4 J_{yy}$ respectively and when $J_\pm=-0.3$, $h=0.2 J_{yy}$ (b). The First Brillouin zone is highlighted in white.}
    \label{fig:sublattice110_2}
\end{figure}

\begin{figure}
    \centering
    \includegraphics[width=\linewidth]{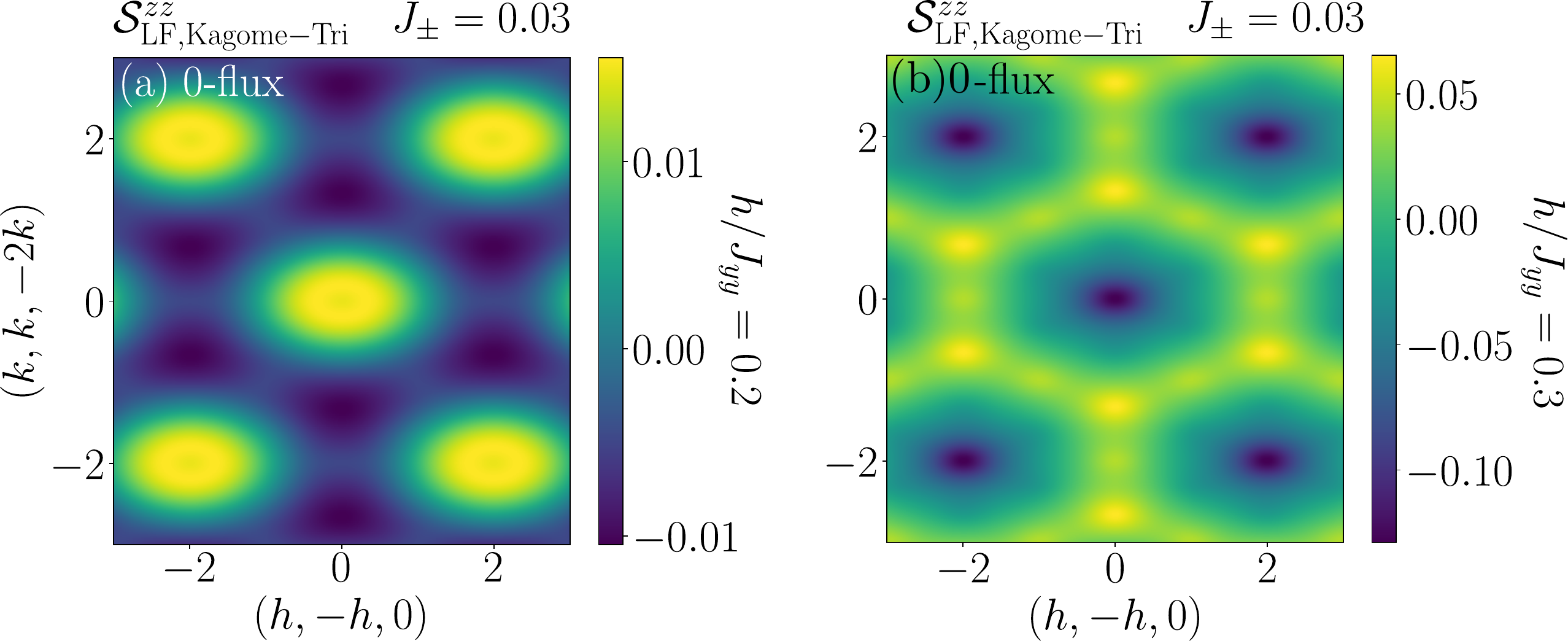}
    \caption{Sublattice SSSF under a $[111]$ field in the local frame for correlations between the Kagome and triangular planes $\mathcal{S}^{zz}_{\mathrm{LF, Kagome-Tri}}=\mathcal{S}^{zz}_{\mathrm{LF,01}}+\mathcal{S}^{zz}_{\mathrm{LF,02}}+\mathcal{S}^{zz}_{\mathrm{LF,03}}$ when $J_\pm=-0.03$ for  (a) $h=0.2J_{yy}$ and (b) $h=0.3 J_{yy}$.} 
    \label{fig:sublattice111}
\end{figure}

\section{Physical Interpretation of Experimental Signatures}

\subsection{Classical Picture}
To conceptually understand the results in the main text, let us first roughly discuss the underlying physics. In the Ising limit where $J_\pm=0$ and $h=0$, states in the degenerate ground state manifold follow the classical 2-in-2-out ice rule in the $\ket{S^y=\pm}$ basis with corresponding algebraically decaying equal-time correlations in real space for $\expval{S^y S^y}$ that lead to the celebrated ``pinch points'' in reciprocal space. Here, we have to recall that we are not probing the correlations associated with the spin ice rules $\expval{S^y S^y}$, but rather $\expval{S^z S^z}$. This is one of the key conceptual challenges in theoretically interpreting neutron scattering results for DO QSI. Hence, when rotating such a classical spin ice state from the $\ket{S^y=\pm}$ to the $\ket{S^z=\pm}$ basis, we end up with an equal superposition of all tetrahedron configurations (i.e., 2-in-2-out, 1-in-3-out, 3-in-1-out, and all-in-all-out) with short-range correlations. The corresponding $\expval{S^z S^z}$ correlation is flat and featureless in momentum space. 

When $J_\pm$ becomes non-zero, the ground state becomes a specific linear superposition of the classical spin ice states that leads to 0-flux ($\pi$-flux) for $J_{\pm}>0$ ($J_{\pm}<0$). Translated to the $\ket{S^z=\pm}$ basis once again, this implies that a ferromagnetic transverse coupling ($J_{\pm}>0$) will favor the ferromagnetic AIAO configurations. In contrast, for $\pi$-flux QSI with $J_{\pm}<0$, the 2-in-2-out configurations are favored. On the other hand, at large magnetic fields $h/J_{yy}$, the states illustrated in Fig.~\ref{fig:pyrochlore}(d)-(f) are energetically preferred. Therefore, the evolution of the neutron scattering signal as we turn on the magnetic field should be indicative of this competition between the weak field configuration described above and the strong field states depicted in Fig.~\ref{fig:pyrochlore}(d)-(f). 

With this picture in mind, we will provide a physical interpretation of the evolution of the SSSF. To gain insights into these results, it will be useful to decompose the signal into contributions from different sublattice pairs by picking the appropriate index in equation~\eqref{eq:SSF}.

\subsection{\texorpdfstring{$\mathbf{B}\parallel[110]$}{\texttwoinferior}}\label{sec:sublatticeSSSF110}

The reason behind the rod-like intensity along $(0,0,l)$ and the emergent pinch point for a field along the [110] direction is due to the increasing polarization of the $\alpha$ chains. To see this, we can isolate the $\alpha$ chain contribution by looking at its sublattice SSSF $\mathcal{S}^{zz}_{03}$ 
 in Fig.~\ref{fig:sublattice110_1}. The intensities in $\mathcal{S}^{zz}_{03}$ increase as we increase the magnetic field, and its shape gives rise to the heightened intensity around $(0,0,l)$. It further yields a ``pinch point" at the zone center that simply comes from the transverse projector $(\hat{\mathbf{z}}_\mu\cdot \mathbf{q})(\hat{\mathbf{z}}_\nu\cdot \mathbf{q})/|\mathbf{q}|^2$ in Eq.~\eqref{eq:szzmunu} and emerges as a result of a pronounced $\alpha$ chain signal. Even though this seems like a universal behavior under a [110] field, the evolution of the 0-flux phase is more nuanced. Recall that, under a weak field, the AIAO configuration is energetically favored for $J_\pm<0$. Indeed, we see that in Fig.~\ref{fig:sublattice110_1}(a) that $\mathcal{S}^{zz}_{03}$ exhibits ferromagnetic signature opposite to that in Fig.~\ref{fig:sublattice110_1}(c). At large enough fields, the antiferromagnetic alignment in $S^z$ of sites 0 and 3, as shown in Fig.~\ref{fig:pyrochlore}(d), is preferred. Therefore, the ferromagnetic AIAO configuration in the weak-field 0-flux state must experience a cross-over which results in the heightened intensities in $ \mathcal{S}^{zz}$ around $(0,0,l)$ in Fig.~\ref{fig:SSSF110}(c).

 On the other hand, we can also isolate the signal coming from the $\beta$ chains. If we only consider the $\beta$ chains, the Brillouin zone center $\Gamma_\beta$ would be located are $(\pm2,\mp2,0)$, as shown in Fig.~\ref{fig:sublattice110_3}, and the high-symmetry points at the zone boundary $K_\beta$ are at $(\pm1,\mp 1,0)$. We see in Fig.~\ref{fig:sublattice110_3} that, when $J_\pm>0$, the local frame SSSF shows $1d$ ferromagnetic signatures with peaks at the $\Gamma_\beta$ points. In contrast, when $J_\pm<0$ we instead see antiferromagnetic signatures with intensities at $K_\beta$ points in Fig.~\ref{fig:sublattice110_3}(b). This dichotomy translates to the snowflake patterns with opposite intensities in $\mathcal{S}^{zz}$ as shown in Fig.~\ref{fig:SSSF110}(a) and (b). One important note is that, even though, in Fig.~\ref{fig:sublattice110_3}(d), we have transitioned from the $\pi$-flux state to the $(0,\pi,\pi,0)$ phase, the intra-$\beta$-chain correlation remains antiferromagnetic. 
 
As the $\alpha$ chains get increasingly polarized, the interchain correlations become much weaker than intra-chain correlations induced by the Zeeman term. As such, we see these stripe-like features develop along the $(0,0,l)$ direction, which is perpendicular to both $\alpha$ and $\beta$ chains. More precisely, these stripe patterns mainly originate from the inter-$\alpha$-chain correlations as they are perpendicular to the scattering plane. We can observe this in Fig.~\ref{fig:sublattice110_2}, where the intensities of the inter-$\alpha$-chain spin correlation $\mathcal{S}^{zz}_{\mathrm{LF}, 00}$ increases as we increase the magnetic field, resulting in a stripe-like pattern at the same location in the global frame.

 Even though the stripe pattern emerges irrespective of $J_\pm$, we see that the positions of the stripes are different for the different phases. Namely, at large fields with $J_\pm=0.03$, the 0-flux state has high intensities at integer values in $(h,-h,0)$. This results from ferromagnetic correlations getting ``stretched" along the $(0,0,l)$ direction due to the aforementioned loss of correlation. This is also true for the $(0,\pi,\pi,0)$-flux phase as shown in Fig.~\ref{fig:sublattice110_2}(c). In contrast, for $\pi$-flux QSI, when $J_\pm=-0.3$ as in Fig.~\ref{fig:synopsis2}(2d), $\mathcal{S}^{zz}$ has extra rod-like intensities stemming from the $L$ points compared to the $(0,\pi,\pi,0)$ case in Fig.~\ref{fig:SSSF110}(d). A close inspection of $\mathcal{S}^{zz}_{\mathrm{LF}, 00}$ shows that this is due to a strong antiferromagnetic inter-$\alpha$-chain correlation of $\pi$-flux phase in Fig.~\ref{fig:sublattice110_2}(b), as evidenced by the rod-like intensities stemming from $X$ and $L$ points. 

\subsection{\texorpdfstring{$\mathbf{B}\parallel[111]$}{\texttwoinferior}}\label{sec:sublatticeSSSF111}
 
As shown before, the Zeeman couplings on the four sublattices for a field parallel to $[111]$ are $-h(\hat{\mathbf{n}}\cdot\hat{\mathbf{z}}_0, \hat{\mathbf{n}}\cdot\hat{\mathbf{z}}_1, \hat{\mathbf{n}}\cdot\hat{\mathbf{z}}_2, \hat{\mathbf{n}}\cdot\hat{\mathbf{z}}_3)=-\frac{h}{3}(3,-1,-1,-1)$. As such, site 0 is strongly coupled to the magnetic field and forms a sparse triangular plane. The other three sites then form a Kagome plane, as shown in Fig.~\ref{fig:pyrochlore}(f). 

At weak fields, 2-in-2-out configurations are favored for $J_\pm<0$. Since sites on the triangular lattice are strongly coupled with the fields, it is always favorable for the $S^z$ components on this plane to align with the [111] field. The 2-in-2-out configuration can still be satisfied if the sites on the Kagome plane follow a 2-in-1-out (1-in-2-out) configuration. This gives rise to the Kagome-ice signature shown in the main text.

At weak fields for $J_\pm>0$, an AIAO configuration is favored where the spins point towards the centers of the down-pointing tetrahedrons such that the spins on sublattice 0 are polarized along the [111] direction. As such, the Kagome plane follows an AIAO configuration. At large enough magnetic fields, all the spins are polarized along the [111] direction as shown in Fig.~\ref{fig:pyrochlore}(f), where the spins on the Kagome planes should point away from the centers of the down-pointing tetrahedrons. This configuration is the opposite of the weak field picture in the Kagome plane, and hence, we should expect a cross-over in this plane. Indeed, if we look at the correlation between Kagome and the triangular plane $\mathcal{S}^{zz}_{\mathrm{Kagome-Tri}}=\mathcal{S}^{zz}_{01}+\mathcal{S}^{zz}_{02}+\mathcal{S}^{zz}_{03}$, we see that in Fig.~\ref{fig:sublattice111}(a)-(b) the intensities flip, signaling that the correlations between site 0 and sites 1,2,3 are now antiferromagnetic. This results in the emergent intensity at the $\Gamma$ point in Fig.~\ref{fig:SSSF111}(c).

\section{Dynamical Spin Structure Factors}
\begin{figure*}
    \centering
    \includegraphics[width=\linewidth]{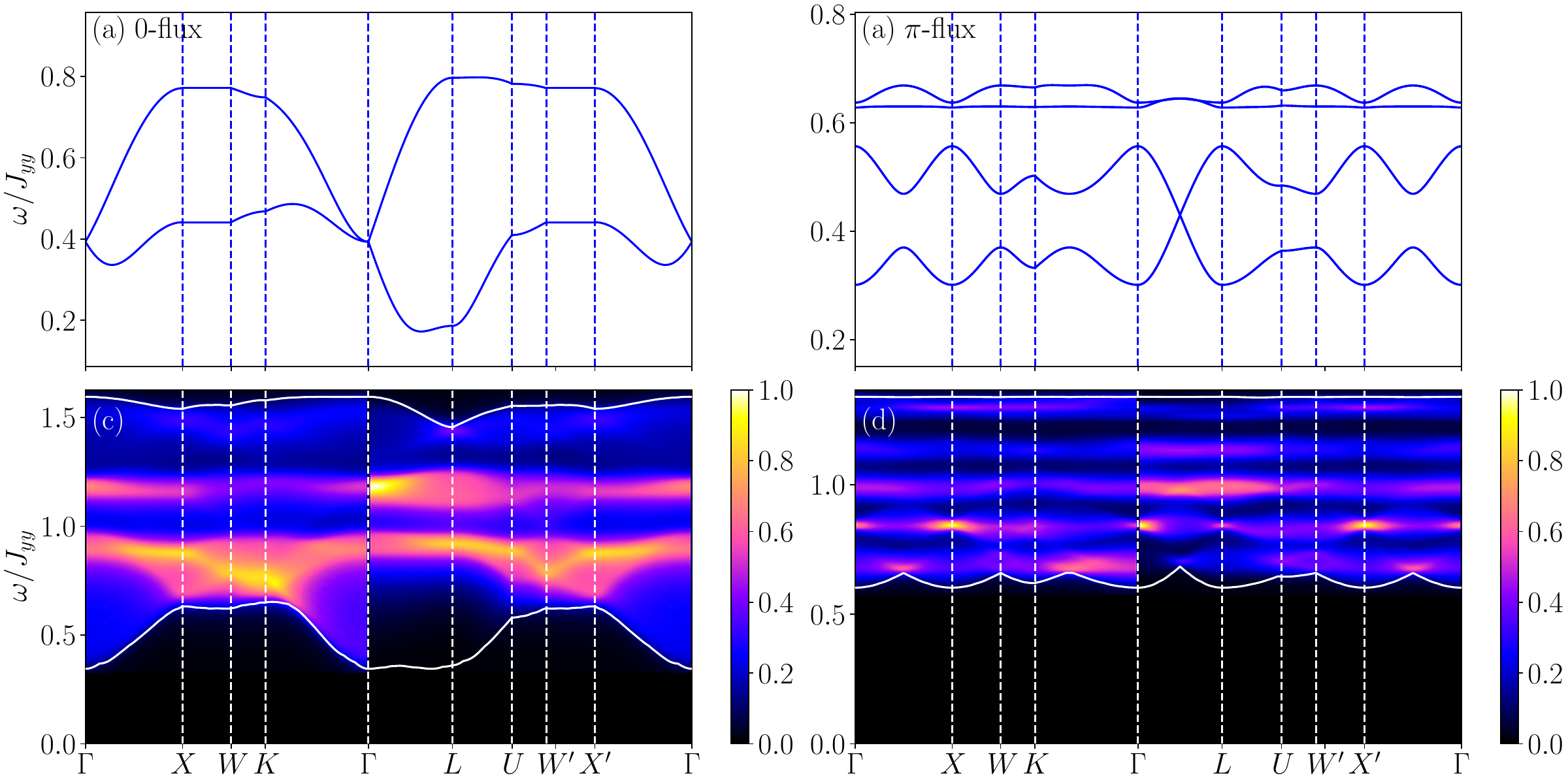}
    \caption{(a)-(b) Spinon dispersions and (c)-(d) dynamical spin structure factor in the global frame under a $[111]$ magnetic field for (a), (c) $J_\pm/J_{yy}=0.03$ and $h/J_{yy}=0.3$, and  (b), (d) $J_\pm/J_{yy}=-0.03$, $h/J_{yy}=0.2$. The upper and lower edges of the two-spinon continuum are denoted by white lines.}
    \label{fig:DSSF111}
\end{figure*}

\begin{figure*}
    \centering
    \includegraphics[width=\linewidth]{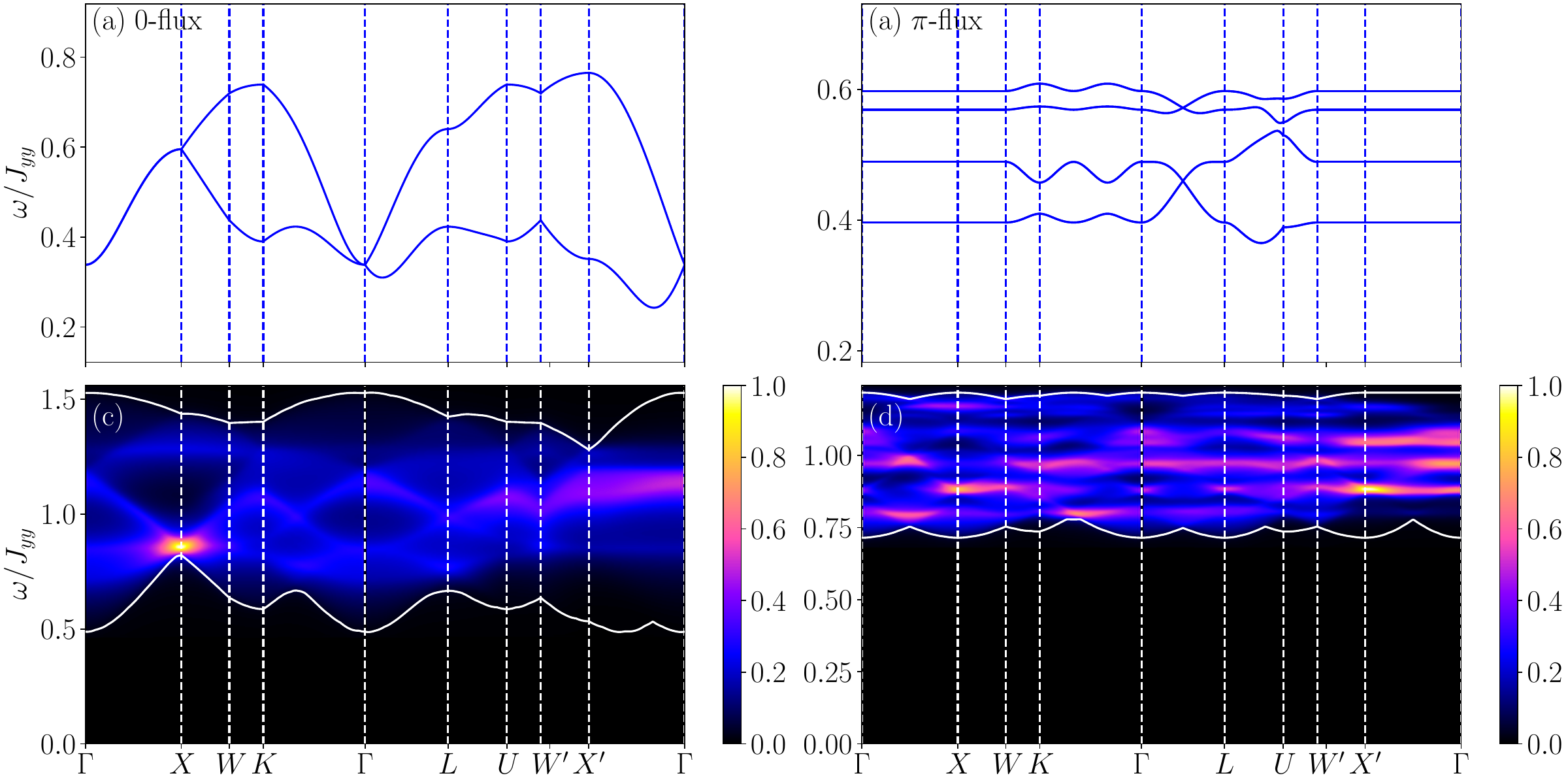}
    \caption{(a)-(b) Spinon dispersions and (c)-(d) dynamical spin structure factor in the global frame under a $[001]$ magnetic field for (a), (c) $J_\pm/J_{yy}=0.03$ and $h/J_{yy}=0.2$, and  (b), (d) $J_\pm/J_{yy}=-0.03$ and $h/J_{yy}=0.1$. The upper and lower edges of the two-spinon continuum are denoted by white lines.}
    \label{fig:DSSF001}
\end{figure*}

We show the spinon dispersion and DSSF under [111] and [001] fields in Fig.~\ref{fig:DSSF111} and \ref{fig:DSSF001}, respectively.

\section{Enhanced Periodicity of the Staggered Flux Phase \texorpdfstring{$(0,\pi,\pi,0)$}{\texttwoinferior}\label{sec:enhancedperiod}}

A signature of the $\pi$-flux phase is an enhanced spectral periodicity that stems from the fractionalization of translation symmetries as highlighted in~\cite{essin2014spectroscopic, chen2017spectral, desrochers2023symmetry}. Here, we comment on what happens to spectral periodicity enhancement in the staggered flux phase $(0,\pi,\pi,0)$. Since only half of the plaquettes are $\pi$-flux and the other half are 0-flux, the (anti)communication rules for the spinon translations are as follows
\begin{subequations}
\begin{align}
    \{T_1^s, T_3^s\}&=0\\
    \{T_2^s, T_3^s\}&=0\\
    [T_1^s, T_2^s] &=0.
\end{align}
\end{subequations}
As such, following the same argument in Ref.~\cite{chen2017spectral}, we construct two-spinon scattering states: 
\begin{subequations}
\begin{align}
    |b\rangle&=T_1^s(1)|a\rangle\\
    |c\rangle&=T_2^s(1)|a\rangle\\
    |d\rangle&=T_3^s(1)|a\rangle
\end{align}
\end{subequations}
by applying single spinon translation on some generic-two spinon scattering state $|a\rangle=|\mathbf{q}_a,z_a\rangle$. Here $\mathbf{q}_a$ is the crystal momentum, and $z_a$ is some remaining quantum number. Let us denote $\mathbf{q}_a=\sum_i q_i \mathbf{G}_i$, where $\mathbf{G}_i$ are the reciprocal space basis vector for FCC systems: $\mathbf{G}_1 = \pi(-1,1,1)$, $\mathbf{G}_2 = \pi(1,-1,1)$, and $\mathbf{G}_3 = \pi(1,1,-1)$ in the cubic coordinates. Lattice translations $T_\mu$ act on the above states as spinon translation on both the the `1' and `2' spinons: $T_\mu\ket{a}=T^s_\mu(1) T^s_\mu(2)\ket{a}$. Acting with translations on $\ket{b}$ yields
\begin{subequations}
\begin{align}
& T_1|b\rangle=T_1^s(1) T_1^s(2) T_1^s(1)|a\rangle=+T_1^s(1)\left[T_1|a\rangle\right] \\
& T_2|b\rangle=T_2^s(1) T_2^s(2) T_1^s(1)|a\rangle=+T_1^s(1)\left[T_2|a\rangle\right] \\
& T_3|b\rangle=T_3^s(1) T_3^s(2) T_1^s(1)|a\rangle=-T_1^s(1)\left[T_3|a\rangle\right].
\end{align} 
\end{subequations}
We can repeat the same for $\ket{c}$ and $\ket{d}$ and obtain
\begin{subequations}
\begin{align}
\mathbf{q}_b &= \mathbf{q}_a + (0,0,1)\\
\mathbf{q}_c &= \mathbf{q}_a + (0,0,1)\\
\mathbf{q}_d &= \mathbf{q}_a + (1,1,0).
\end{align}
\end{subequations}
This implies enhanced spectral periodicity at the points above since $\ket{a}$, $\ket{b}$, $\ket{c}$, and $\ket{d}$ have the same energy and the same quantum number. Converting to the cubic coordinate systems, we will find enhanced spectral periodicity at $\mathbf{q}$, $\mathbf{q}+\pi(1,1,-1)$ and $\mathbf{q}+2\pi(0,0,1)$. The distinct spectral periodicity can potentially be used as a useful way to distinguish experimentally (or numerically) 0-flux QSI, $\pi$-flux QSI, and the $(0,\pi,\pi,0)$ state.

\bibliography{ref}

\end{document}